\documentclass[sigconf]{acmart}
\usepackage{epsfig,endnotes}
\usepackage{multirow}

\usepackage{graphicx}
\usepackage{subfig}
\usepackage{grffile}

\newcounter{subcopyrightbox@save}
\usepackage{caption}
\usepackage{color, url}
\usepackage{xspace} 

\usepackage{mathrsfs}

\usepackage{amsmath}
\usepackage{epstopdf}
\usepackage{balance}
\usepackage{bm}
\usepackage{makecell}
\usepackage{cleveref}

\usepackage{algorithm}
\usepackage{algorithmic}

\newcommand\SHF[1]{{#1}}

\copyrightyear{2021}
\acmYear{2021}
\setcopyright{acmcopyright}
\acmConference[CCS '21] {2021 ACM Conference on Computer and Communications Security}{November 15--19, 2021}{Coex, Seoul, South Korea}

\acmPrice{15.00}
\acmISBN{xxx-1-xxxx-xxxx-8/21/06}
\acmDOI{10.1145/xxxxxxx.xxxxxxx}

\newcommand{\myparatight}[1]{\smallskip\noindent{\bf {#1}:}~}

\newenvironment{packeditemize}{\begin{list}{$\bullet$}{\setlength{\itemsep}{1pt}\addtolength{\labelwidth}{0pt}\setlength{\leftmargin}{\labelwidth}\setlength{\listparindent}{\parindent}\setlength{\parsep}{0pt}\setlength{\topsep}{0pt}}}{\end{list}}

\graphicspath{ {figs/} }

\begin{document}

\date{}
\fancyhf{}

\title{\SHF{EncoderMI: Membership Inference against Pre-trained \\ Encoders in Contrastive Learning}}

\author{Hongbin Liu$^*$}
\affiliation{%
  \institution{Duke University}
}
\email{hongbin.liu@duke.edu}

\author{Jinyuan Jia$^*$}
\affiliation{%
  \institution{Duke University}
}
\email{jinyuan.jia@duke.edu}

\author{Wenjie Qu$^\dagger$}
\affiliation{%
  \institution{Huazhong University of Science and Technology}
}
\email{wenjiequ@hust.edu.cn}

\author{Neil Zhenqiang Gong}
\affiliation{%
  \institution{Duke University}
}
\email{neil.gong@duke.edu}

\thanks{$^*$The first two authors made equal contributions.}
\thanks{$^\dagger$Wenjie Qu performed this research when he was a remote intern in Gong's group.}

\begin{abstract}
\SHF{Given a set of unlabeled images or (image, text) pairs, \emph{contrastive learning} aims to pre-train an \emph{image encoder} that can be used as a feature extractor for many downstream tasks.  In this work, we propose \SHF{\emph{EncoderMI}}, the first \emph{membership inference} method against image encoders pre-trained by contrastive learning.} In particular, given an input and a black-box access to an image encoder,  \SHF{EncoderMI} aims to infer whether the input is in the training dataset of the image encoder. \SHF{EncoderMI can be used 1) by a data owner to audit whether its (public) data was used to pre-train an image encoder without its authorization or 2) by an attacker to compromise privacy of the training data when it is private/sensitive.}   
Our \SHF{EncoderMI} exploits the \emph{overfitting} of the image encoder towards its training data. In particular, an overfitted image encoder is more likely to output more (or less) similar feature vectors for two augmented versions of an input in (or not in) its training dataset. We evaluate \SHF{EncoderMI} on image encoders pre-trained on multiple datasets by ourselves as well as the Contrastive Language-Image Pre-training (CLIP) image encoder, which is pre-trained on 400 million (image, text) pairs collected from the Internet and released by OpenAI. Our results show that \SHF{EncoderMI} can achieve high \emph{accuracy}, \emph{precision}, and \emph{recall}. We also explore a countermeasure against \SHF{EncoderMI} via preventing overfitting through early stopping. \SHF{Our  results show that it achieves trade-offs between accuracy of \SHF{EncoderMI} and utility of the image encoder, i.e., it can reduce the accuracy of \SHF{EncoderMI}, but it also incurs classification accuracy loss of the downstream classifiers built based on the image encoder.} 

\end{abstract}






\maketitle
\section{Introduction}
\SHF{\emph{Contrastive learning}~\cite{hadsell2006dimensionality,oord2018representation,he2020momentum,chen2020simple,radford2021learning} is a promising approach for general-purpose AI. In particular, given an unlabeled dataset (called \emph{pre-training dataset}) of images or (image, text) pairs, contrastive learning pre-trains an \emph{image encoder} that can be used as a feature extractor for many downstream tasks.} Given the image encoder, the downstream tasks require only a small amount of or no labeled training data. The pre-training of encoders, however, usually consumes a lot of data and computation resources. Therefore, typically, a powerful encoder provider (e.g., OpenAI, Google) pre-trains encoders and then provides service to downstream customers (e.g., less resourceful organizations, end users).  

Existing studies~\cite{he2020momentum,chen2020simple,radford2021learning} on \SHF{contrastive learning} mainly focus on how to train a better image encoder such that it can achieve better performance on the downstream tasks. \SHF{The security and privacy of contrastive learning}, however, is largely unexplored. In this work, we perform the first systematic study on \emph{membership inference} against image encoders pre-trained by \SHF{contrastive learning}. In particular, we aim to infer whether an input image is in the pre-training dataset of an image encoder. An input is called a \emph{member} (or \emph{non-member}) of an image encoder if it is in (or not in) the pre-training dataset of the image encoder. 

\SHF{Membership inference in contrastive learning has two important applications.  Suppose data owners make their images public on the Internet, e.g., on Twitter. An AI company (e.g., OpenAI)  collects and uses the public data to pre-train and monetize image encoders without the data owners' authorization. Such practices may violate the data owners' data security. For instance, Twitter asked Clearview to stop taking public images from its website for model training~\cite{twitter_stop_url}; and FTC requires Ever to delete  models trained on unauthorized user data~\cite{FTC_settlement}. The first application of  membership inference is that a data owner can use a membership inference method to audit whether his/her (public) data was used to pre-train image encoders without his/her authorization, though the membership inference result may not have formal guarantees.  The second application of  membership inference is that an attacker can use it to compromise  privacy of the pre-training data when it is private/sensitive. For instance, hospitals may collaboratively use contrastive learning to pre-train image encoders that can be shared across hospitals to solve various downstream healthcare tasks, e.g.,  lung CT image based COVID-19 testing~\cite{fung2021self} and skin disease prediction. In such cases, the pre-training data may include sensitive medical images and one hospital may infer other hospitals' sensitive members of the image encoder. }

Existing membership inference \SHF{methods}~\cite{shokri2017membership,salem2018ml,nasr2018machine,nasr2019comprehensive,jia2019memguard,song2019privacy} are mainly designed for classifiers. For example, given the confidence score vector outputted by a classifier for an input, they aim to infer whether the input is in the training dataset of the classifier. The idea of existing membership inference \SHF{methods}~\cite{shokri2017membership,salem2018ml} is to exploit the \emph{overfitting} of the classifier. For instance, the confidence score vectors of members and non-members of a classifier are statistically distinguishable. Therefore, the confidence score vector outputted by the classifier for an input can capture whether the input is a member of the classifier. However, given an input, an image encoder outputs a feature vector for it. The feature vector itself does not capture the overfitting of the image encoder on the input. As shown by our experimental results in Section~\ref{exp-results}, existing membership inference \SHF{methods} for classifiers are close to random guessing when generalized to infer the members of an image encoder. 

\myparatight{Our work} In this work, we propose \SHF{EncoderMI}, the first \SHF{membership inference method against contrastive learning}. 

\emph{Threat model.} \SHF{We call an entity (e.g., a data owner, an attacker) who performs membership inference an \emph{inferrer}.} We assume an \SHF{inferrer} has a black-box access to a pre-trained image encoder (called \emph{target encoder}), which is the most difficult and general scenario. The \SHF{inferrer} aims to infer whether an input is in the pre-training dataset of the target encoder. The pre-training of an image encoder relies on three key dimensions: \emph{pre-training data distribution}, \emph{encoder architecture}, and \emph{training algorithm}. In other words, we have three dimensions of background knowledge. \SHF{The inferrer} may or may not know each of them. Therefore, we have eight different types of background knowledge for \SHF{the inferrer}  in total.  
In our \SHF{methods}, we assume the \SHF{inferrer} has a \emph{shadow dataset}. In particular, the shadow dataset could have the same distribution as the pre-training data distribution if the \SHF{inferrer} knows it. Otherwise, we assume the shadow dataset has a different distribution from the pre-training dataset. Moreover, if the \SHF{inferrer} does not know the encoder architecture (or training algorithm), we consider the \SHF{inferrer} can assume one and perform membership inference based on the assumed one. 

\emph{Our \SHF{EncoderMI}.}
An important module in \SHF{contrastive learning} is \emph{data augmentation}. Roughly speaking, given an input, the data augmentation module creates another random input (called \emph{augmented input}) via applying a sequence of random operations (e.g., random grayscale, random resized crop) to the input. We observe that \SHF{contrastive learning} essentially aims to pre-train an image encoder such that it outputs similar feature vectors for two augmented inputs created from the same input.  \SHF{EncoderMI} is based on this observation. Specifically, when an image encoder is \emph{overfitted} to its pre-training dataset, it may output more (or less) similar feature vectors for augmented inputs that are created from an input in (or not in) the pre-training dataset.  In \SHF{EncoderMI}, \SHF{an inferrer} builds a binary classifier (called \emph{\SHF{inference classifier}}) to predict whether an input is a member of the target encoder. Roughly speaking, our \SHF{inference classifier} predicts an input to be a member of the target encoder if the target encoder produces similar feature vectors for the augmented inputs created from the input. Next, we discuss how to build \SHF{inference classifiers}.

Given a shadow dataset, we first split it into two subsets, namely, \emph{shadow member dataset} and \emph{shadow non-member dataset}. Then, we \SHF{pre-train an encoder (called \emph{shadow encoder})} using the shadow member dataset based on the background knowledge of the \SHF{inferrer} (e.g., the \SHF{inferrer} can adopt the same encoder architecture and training algorithm used to pre-train the target encoder if he/she knows them). Given the shadow encoder and the shadow dataset, we extract \emph{membership features} for each input in the shadow dataset. In particular, given an input in the shadow dataset, we first create $n$ augmented inputs via the data augmentation module of the training algorithm used to train the shadow encoder, then use the shadow encoder to produce a feature vector for each augmented input, and finally compute the set of $n\cdot (n-1)/2$ pairwise similarity scores between the $n$ feature vectors using a similarity metric as the membership features for the input. Given these membership features, we construct an \emph{\SHF{inference training dataset}} via labeling the membership features as ``member'' ( or ``non-member'') if they are created from an input that is in the shadow member (or non-member) dataset. Given the \SHF{inference training dataset}, we build an \SHF{inference classifier} to infer the members of the target encoder.  We consider three types of classifiers: \emph{vector-based classifier}, \emph{set-based classifier}, and \emph{threshold-based classifier}. Given an input and a black-box access to the target encoder, we first extract membership features for the input and then use an \SHF{inference classifier} to predict whether the input is a member of the target encoder.  

\emph{Evaluation.}
To evaluate \SHF{EncoderMI}, we first conduct experiments on CIFAR10, STL10, and Tiny-ImageNet datasets via pre-training image encoders by ourselves. Our experimental results show that \SHF{EncoderMI} can achieve high \emph{accuracy}, \emph{precision}, and \emph{recall} under all the eight different types of background knowledge. For instance, our vector-based \SHF{inference classifier} can achieve 88.7\% – 96.5\% accuracy on Tiny-ImageNet under the eight types of background knowledge. Moreover, \SHF{EncoderMI} can achieve higher accuracy as the \SHF{inferrer} has access to more background knowledge. We also apply \SHF{EncoderMI} to infer members of the CLIP's image encoder released by OpenAI~\cite{radford2021learning}. In particular, we collect some \emph{potential members} and ground truth non-members of the CLIP's image encoder from Google image search and Flickr. Our results show that \SHF{EncoderMI} is effective even if the \SHF{inferrer} does not know the pre-training data distribution, the encoder architecture, and the training algorithm of the CLIP's image encoder. 

\emph{Countermeasure.} \SHF{When a data owner uses EncoderMI to audit data misuse, an encoder provider may adopt a countermeasure against EncoderMI to evade auditing. When an attacker uses EncoderMI to compromise pre-training data privacy, a countermeasure  can be adopted to enhance privacy.}  
As  \SHF{EncoderMI} exploits the overfitting of the target encoder on its training data, we can leverage countermeasures that prevent overfitting. In particular, we generalize early stopping~\cite{song2020systematic}, a state-of-the-art overfitting-prevention-based countermeasure against membership inference to classifiers,  to mitigate   \SHF{membership inference to pre-trained encoders}. Roughly speaking, the idea of early stopping is to train the target encoder with less number of epochs to prevent overfitting. Our results show that it achieves  trade-offs \SHF{between the accuracy of EncoderMI and the utility of the target encoder.} More specifically, it can reduce the accuracy of our \SHF{EncoderMI}, but it also incurs  classification accuracy loss of the downstream classifiers built based on the target encoder.

In summary, we make the following contributions in this work: 
\begin{packeditemize}
    \item We propose \SHF{EncoderMI}, the first membership inference \SHF{method} against \SHF{contrastive} learning. 
    \item We conduct extensive experiments to evaluate our \SHF{EncoderMI} on CIFAR10, STL10, and Tiny-ImageNet datasets. Moreover, we apply \SHF{EncoderMI} to CLIP's image encoder. 
    \item We evaluate an early stopping based countermeasure against  \SHF{EncoderMI}. Our results show that it achieves trade-offs \SHF{between accuracy of EncoderMI and utility of the encoder}.
\end{packeditemize}
\section{Background on Contrastive Learning}
\label{sec:background}

\SHF{Given a large amount of unlabeled images or (image, text) pairs (called \emph{pre-training dataset}), contrastive learning aims to pre-train a neural network (called \emph{image encoder}) that can be used as a feature extractor for many downstream tasks (e.g., image classification). Given an input image, the pre-trained image encoder outputs a \emph{feature vector} for it. }

\subsection{Pre-training an Encoder} 
 An essential module in \SHF{contrastive learning} is \emph{data augmentation}. Given an input image,  the  data augmentation module can create another random input (called \emph{augmented input}) by a sequence of random operations such as random grayscale, random resized crop, etc.. An augmented input and the original input have the same size. Moreover, we can create multiple augmented inputs for each input using the data augmentation module. 
 Roughly speaking, the idea of contrastive learning is to pre-train an image encoder such that it outputs similar (or dissimilar) feature vectors for two augmented inputs created from the same (or different) input(s). Contrastive learning formulates such similarity as a \emph{contrastive loss}, which an image encoder is trained to minimize. Next, we introduce three popular contrastive learning algorithms, i.e., MoCo~\cite{he2020momentum}, SimCLR~\cite{chen2020simple}, and CLIP~\cite{radford2021learning}, to further illustrate the idea of contrastive learning.

\noindent
{\bf MoCo~\cite{he2020momentum}:} MoCo pre-trains an image encoder on unlabeled images.  There are three major modules in MoCo: an \emph{image encoder} (denoted as $h$), a \emph{momentum encoder} (denoted as $h_m$), and a \emph{dictionary} (denoted as $\Gamma$). The image encoder outputs a feature vector for an input or an augmented input. The momentum encoder has the same  architecture with the image encoder, but is updated much more slowly compared with the image encoder. Given an input or an augmented input, the momentum encoder also outputs a vector for it. To distinguish with feature vector, we call it \emph{key vector}. The \emph{dictionary} module maintains a queue of key vectors outputted by the momentum encoder for augmented inputs created from inputs in previous several mini-batches. Moreover, the dictionary is dynamically updated during the pre-training of the image encoder. 

Given a mini-batch of $N$ inputs, MoCo creates two augmented inputs for each input in the mini-batch. The two augmented inputs are respectively passed to the image encoder and the momentum encoder. For simplicity, we use $\mathbf{u}_i$ and $\mathbf{u}_j$ to denote these two augmented inputs. Given the two augmented inputs, the image encoder $h$, the momentum encoder $h_m$, and the dictionary $\Gamma$, MoCo defines a contrastive loss  as follows:
\begin{align}
    &\ell(\mathbf{u}_i) \nonumber\\
    = &- \log(\frac{exp(sim(h(\mathbf{u}_i), h_m(\mathbf{u}_j))/\tau)}{exp(sim(h(\mathbf{u}_i), h_m(\mathbf{u}_j))/\tau) +\sum\limits_{\mathbf{z} \in \Gamma} exp(sim(h(\mathbf{u}_i), \mathbf{z})/\tau)}),
\end{align}
where $exp$ is the natural exponential function, $sim$ computes cosine similarity between two vectors, and $\tau$ represents a temperature parameter. The final contrastive loss is summed over the contrastive loss $\ell(\mathbf{u}_i)$ of the $N$ augmented inputs (i.e., $u_i$'s) corresponding to the $N$ inputs. MoCo pre-trains the image encoder via minimizing the final contrastive loss. Finally, the $N$ key vectors (i.e., $h_m(\mathbf{u}_j)$'s) outputted by the momentum encoder for the $N$ augmented inputs (i.e., $u_j$'s) are enqueued to the dictionary while the $N$ key vectors from the ``oldest'' mini-batch are dequeued.

\noindent
{\bf SimCLR~\cite{chen2020simple}:} Similar to MoCo~\cite{he2020momentum}, SimCLR also tries to pre-train an image encoder on unlabeled images. Given a mini-batch of $N$ inputs, SimCLR creates two augmented inputs for each input in the mini-batch via data augmentation. Given $2 \cdot N$ augmented inputs (denoted as $\{\mathbf{u}_1, \mathbf{u}_2, \cdots, \mathbf{u}_{2\cdot N}\}$), SimCLR aims to pre-train the image encoder such that it outputs similar (or dissimilar) feature vectors for two augmented inputs that are created from the same (or different) input(s). 
Formally, given a pair $(\mathbf{u}_i, \mathbf{u}_j)$ of augmented inputs created from the same input, the contrastive loss is defined as follows:
\begin{align}
    \ell_{ij} = - \log(\frac{exp(sim(g(h(\mathbf{u}_i)), g( h(\mathbf{u}_j)))/\tau)}{\sum_{k=1}^{2\cdot N}\mathbb{I}(k \neq i)\cdot exp(sim(g( h(\mathbf{u}_i)), g( h(\mathbf{u}_k)))/\tau)}),
\end{align}
where $\mathbb{I}$ is an indicator function, $exp$ is the natural exponential function, $sim$ computes cosine similarity between two vectors, $h$ is the image encoder, $g$ is the projection head, and $\tau$ is a temperature parameter. The final contrastive loss is summed over the contrastive loss $\ell_{ij}$ of all $2 \cdot N$ pairs of augmented inputs, where each input corresponds to two pairs of augmented inputs $(\mathbf{u}_i, \mathbf{u}_j)$ and $(\mathbf{u}_j, \mathbf{u}_i)$.  SimCLR pre-trains the image encoder via minimizing the final contrastive loss.

\noindent
{\bf CLIP~\cite{radford2021learning}:} CLIP jointly pre-trains an image encoder and a text encoder on unlabeled (image, text) pairs. In particular, the text encoder takes a text as input and outputs a {feature vector} for it. Given a mini-batch of $N$ (image, text) pairs, CLIP creates an augmented input image from each input image. For each augmented input image, CLIP forms a correct (image, text) pair using the augmented input image and the text that originally pairs with the input image from which the augmented input image is created, and CLIP forms $(N-1)$ incorrect pairs using the augmented input image and the remaining $(N-1)$ texts. Therefore, there are  $N$ correct pairs and $N\cdot (N-1)$ incorrect pairs in total. CLIP jointly pre-trains an image encoder and a text encoder such that, for a correct (or incorrect) pair of (image, text), the feature vector outputted by the image encoder for the augmented input image is similar (or dissimilar) to the feature vector outputted by the text encoder for the text.  

\noindent
{\bf Observation:} We observe that these contrastive learning algorithms try to pre-train an image encoder that outputs similar feature vectors for two augmented inputs that are created from the same input. Specifically, we can have this observation for  MoCo~\cite{he2020momentum} and SimCLR~\cite{chen2020simple} based on the definitions of their contrastive losses. 
For CLIP, given an (image, text) pair,  the feature vector outputted by the image encoder for an augmented version of the image is similar to the feature vector outputted by the text encoder for the text. Therefore, the feature vectors outputted by the image encoder for two augmented inputs created from the input image are similar since both of them are similar to the feature vector outputted by the text encoder for the given text. As we will discuss in Section~\ref{sec:attack}, our EncoderMI leverages this observation to infer members of an image encoder's pre-training dataset.

\subsection{Training Downstream Classifiers}
The image encoder can be used as a feature extractor for many downstream tasks. We consider the downstream task to be image classification in this work.
In particular, suppose we have a labeled dataset (called \emph{downstream dataset}). We first use the image encoder to extract feature vectors for inputs in the downstream dataset. Then, we follow the standard supervised learning to train a classifier (called \emph{downstream classifier}) using the extracted feature vectors as well as the corresponding labels. Given a testing input from the downstream task, we first use the image encoder to extract the feature vector for it and then use the downstream classifier to predict a label for the extracted feature vector. The predicted label is viewed as the prediction result for the testing input.

\section{Problem Formulation}

\subsection{Threat Model}
\noindent
{\bf \SHF{Inferrer}'s goal:} Given an input image $\mathbf{x}$, an \SHF{inferrer} aims to infer whether it is in the pre-training dataset of an image encoder (called \emph{target encoder}). 
We call an input a \emph{member} of the target encoder if the input is in its pre-training dataset, otherwise we call the input a \emph{non-member}. 
The \SHF{inferrer} aims to achieve high  accuracy at inferring the members/non-members of the target encoder.

\noindent
{\bf \SHF{Inferrer}'s background knowledge:}  We consider an \SHF{inferrer} has a \emph{black-box} access to the target encoder.  We note that this is the most difficult and general scenario for the \SHF{inferrer}. A typical application scenario is that the encoder provider pre-trains an encoder and then provides an API to downstream customers. 
The pre-training of an encoder depends on three key dimensions, i.e., \emph{pre-training data distribution}, \emph{encoder architecture}, and \emph{training algorithm} (e.g., MoCo, SimCLR). 
Therefore, we characterize the \SHF{inferrer}'s background knowledge along these three dimensions.
\begin{packeditemize}
    \item {\bf Pre-training data distribution.} This background knowledge characterizes whether the \SHF{inferrer} knows the distribution of the pre-training dataset of the target encoder. In particular, if the \SHF{inferrer} knows the distribution, we assume he/she has access to a \emph{shadow dataset} that has the same distribution as the pre-training dataset. Otherwise, we assume the \SHF{inferrer} has access to a shadow dataset that has a different distribution from the pre-training dataset. Note that, in both cases, we consider the shadow dataset does not have overlap with the pre-training dataset.
    For simplicity, we use $\mathcal{P}$ to denote this dimension of background knowledge.

    \item {\bf Encoder architecture.} The \SHF{inferrer} may or may not know the architecture of the target encoder. When the \SHF{inferrer} does not know the target-encoder architecture, the \SHF{inferrer} can assume one and perform \SHF{membership inference} based on the assumed one. For instance, when the target encoder uses ResNet architecture, the \SHF{inferrer} may assume VGG  architecture when performing membership inference. We use $\mathcal{E}$ to denote this dimension of background knowledge.

    \item {\bf Training algorithm.} This dimension characterizes whether the \SHF{inferrer} knows the \SHF{contrastive}  learning algorithm used to train the target encoder. When the \SHF{inferrer} does not know the training algorithm, the \SHF{inferrer} can  perform membership inference based on an assumed one. 
    For instance, when the training algorithm of the target encoder is MoCo~\cite{he2020momentum}, the \SHF{inferrer} may perform \SHF{membership inference} by assuming the training algorithm is SimCLR~\cite{chen2020simple}. We use $\mathcal{T}$ to denote this dimension of background knowledge. 
    
\end{packeditemize}

We use a triplet $\mathcal{B} = (\mathcal{P},\mathcal{E},\mathcal{T})$ to denote the three dimensions of the \SHF{inferrer}'s background knowledge. Each dimension in $\mathcal{B}$ can be ``yes'' or ``no'', where a dimension is ``yes'' ( or ``no'') when the corresponding dimension of background knowledge  is available (or unavailable) to the \SHF{inferrer}. Therefore, we have eight different types of background knowledge in total. 
For instance, the \SHF{inferrer} knows the pre-training data distribution, architecture of the target encoder, and/or training algorithm when the encoder provider makes them public to increase transparency and trust.  

\noindent
{\bf \SHF{Inferrer}'s capability:} An \SHF{inferrer} can query the target encoder for the feature vector of any input or an augmented input.

\subsection{Membership Inference}
Given the \SHF{inferrer}'s goal, background knowledge, and capability, we define our membership inference  against \SHF{contrastive} learning as follows: 

\begin{definition}[Membership Inference against \SHF{Contrastive} Learning]
Given a black-box access to a target encoder, the background knowledge $\mathcal{B}=(\mathcal{P},\mathcal{E},\mathcal{T})$, and an input,  membership inference aims to infer whether the input is in the pre-training dataset of the target encoder. 
\end{definition}

\section{Our Method}
\label{sec:attack}

\subsection{Overview}

Recall that a target encoder is trained to output similar feature vectors for the augmented versions of an input in the pre-training dataset. 
Our \SHF{EncoderMI} is based on this observation. Specifically, when an  encoder is \emph{overfitted} to its pre-training dataset, the  encoder may output more (or less) similar feature vectors for augmented inputs created from an input in (or not in) the pre-training dataset. Therefore,  
our \SHF{EncoderMI} infers an input to be a member of the target encoder if the target encoder produces similar feature vectors for the augmented versions of the input. Specifically, in \SHF{EncoderMI}, an \SHF{inferrer} builds a binary classifier (called \emph{\SHF{inference} classifier}), which predicts member/non-member for an input based on some features we create for the input. To distinguish with the feature vectors produced by the target encoder, we call the features used by the \SHF{inference} classifier \emph{membership features}. Our membership features of an input are based on the similarity scores between the feature vectors of the input's augmented versions produced by the target encoder.  Building our \SHF{inference} classifier requires a training dataset (called \emph{\SHF{inference} training dataset}) which consists of known members and non-members. To construct an \SHF{inference} training dataset, we split the \SHF{inferrer}'s shadow dataset into  two subsets, which we call \emph{shadow member dataset} and \emph{shadow non-member dataset}, respectively.  Then, the \SHF{inferrer} pre-trains an encoder (called \emph{shadow encoder}) using the shadow member dataset. We construct an \SHF{inference} training dataset based on the shadow encoder and shadow dataset. Specifically, each input in  the shadow member (or non-member) dataset is a member (or non-member) of the shadow encoder, and we create membership features for each input in the shadow dataset based on the shadow encoder. After building an \SHF{inference} classifier based on the \SHF{inference} training dataset, we apply it to infer members/non-members of the target encoder.

\subsection{Building \SHF{Inference} Classifiers}
We first introduce how to train a shadow encoder on a shadow dataset. Then, we discuss how to extract membership features for an input. Finally, we discuss how to 
construct an \SHF{inference} training dataset based on the shadow encoder and the shadow dataset, and given the constructed \SHF{inference} training dataset, we discuss how to build \SHF{inference} classifiers.

\noindent
{\bf Training a shadow encoder:} The first step of our \SHF{EncoderMI}  is to train a shadow encoder whose ground truth members/non-members are known to the \SHF{inferrer}. For simplicity, we use $\tilde{h}$ to denote the shadow encoder. In particular,  the \SHF{inferrer} 
splits its shadow dataset $\mathcal{D}_s$ into two non-overlapping subsets: shadow member dataset (denoted as $\mathcal{D}_s^{m}$) and shadow non-member dataset (denoted as $\mathcal{D}_s^{nm}$). Then, the \SHF{inferrer}  pre-trains a shadow encoder using the shadow member dataset $\mathcal{D}_s^{m}$. If the \SHF{inferrer} has access to the target encoder's architecture (or training algorithm), the \SHF{inferrer} uses the same architecture (or training algorithm) for the shadow encoder, otherwise the \SHF{inferrer} assumes an architecture (or training algorithm) for the shadow encoder.  
We note that each input in the shadow member (or non-member) dataset is a member (or non-member) of the shadow encoder. 

\noindent
{\bf Extracting membership features:} For each input in the shadow dataset, we extract its membership features based on the shadow encoder $\tilde{h}$. 
Our membership features are based on the key observation that an encoder (e.g., target encoder, shadow encoder) pre-trained by \SHF{contrastive} learning produces similar feature vectors for augmented versions of an input in the encoder's pre-training dataset.  Therefore, given an input $\mathbf{x}$, we first create $n$ augmented inputs using the data augmentation module $\mathcal{A}$ of the training algorithm used to pre-train the shadow encoder. We denote the $n$ augmented inputs as $\mathbf{x}^1, \mathbf{x}^2, \cdots, \mathbf{x}^{n}$. Then, 
 we use the shadow encoder to produce a feature vector for each augmented input. We denote by $\tilde{h}(\mathbf{x}^i)$ the feature vector produced by the shadow encoder $\tilde{h}$ for the augmented input $\mathbf{x}^i$, where $i=1,2,\cdots,n$. Our \emph{membership features} for the input $\mathbf{x}$ consist of the set of pairwise similarity scores between the $n$ feature vectors. Formally, we have:
\begin{align}
    \mathcal{M}(\mathbf{x}, \tilde{h}) = \{S(\tilde{h}(\mathbf{x}^i), \tilde{h}(\mathbf{x}^j))| i \in [1, n], j \in [1,n], j > i\}, 
\end{align}
where $\mathcal{M}(\mathbf{x}, \tilde{h})$ is our membership features for an input $\mathbf{x}$ based on encoder $\tilde{h}$,   and $S(\cdot, \cdot)$ measures the similarity between two feature vectors (e.g., $S(\cdot, \cdot)$ can be cosine similarity). Note that we omit the explicit dependency of $\mathcal{M}(\mathbf{x},\tilde{h})$ on $S$, $\mathcal{A}$, and $n$ for simplicity.
There are $n \cdot (n-1)/2$ similarity scores in $\mathcal{M}(\mathbf{x}, \tilde{h})$ and they tend to be large if the input $\mathbf{x}$ is a member of the shadow encoder $\tilde{h}$.

\myparatight{Constructing an \SHF{inference} training dataset}
Given the shadow member dataset $\mathcal{D}_s^m$, the shadow non-member dataset $\mathcal{D}_s^{nm}$, and the shadow encoder $\tilde{h}$, we construct an \SHF{inference} training dataset, which is used to build an \SHF{inference} classifier. In particular, given an input $\mathbf{x} \in \mathcal{D}_s^m$, we  extract its membership features $\mathcal{M}(\mathbf{x}, \tilde{h})$ and assign a label $1$ to it; and given an input $\mathbf{x} \in \mathcal{D}_s^{nm}$, we  extract its membership features $\mathcal{M}(\mathbf{x}, \tilde{h})$ and assign a label $0$ to it, where the label $1$ represents ``member" and the label $0$ represents ``non-member". 
Formally, our \SHF{inference} training dataset (denoted as $\mathcal{E}$) is as follows:
\begin{align}
    \mathcal{E} = &\{(\mathcal{M}(\mathbf{x}, \tilde{h}), 1)|\mathbf{x} \in \mathcal{D}_s^{m}\} \cup 
   \{(\mathcal{M}(\mathbf{x}, \tilde{h}), 0)|\mathbf{x} \in \mathcal{D}_s^{nm}\}.
\end{align}

\noindent
{\bf Building \SHF{inference} classifiers:} Given the \SHF{inference} training dataset $\mathcal{E}$, we build a  binary \SHF{inference} classifier. We consider three types of classifiers, i.e., \emph{vector-based classifier}, \emph{set-based classifier}, and \emph{threshold-based classifier}. These classifiers use the membership features differently. Next, we discuss them one by one. 

 \begin{packeditemize}
     \item \emph{\bf Vector-based classifier (\SHF{EncoderMI}-V).} In a vector-based classifier, we transform the set of membership features $\mathcal{M}(\mathbf{x}, \tilde{h})$ of an input into a vector. Specifically, we rank the $n \cdot (n-1)/2$ similarity scores in $\mathcal{M}(\mathbf{x}, \tilde{h})$ in a descending order. We apply the ranking operation to the membership features of each input in the \SHF{inference} training dataset $\mathcal{E}$. Then, we train a vector-based classifier (e.g., a fully connected neural network) on $\mathcal{E}$ following the standard supervised learning procedure.  
    We use $f_v$ to denote the vector-based classifier. Moreover, we use \emph{\SHF{EncoderMI}-V} to denote this \SHF{method}.

     \item {\bf Set-based classifier (\SHF{EncoderMI}-S).} In a set-based classifier, we directly operate on the set of membership features $\mathcal{M}(\mathbf{x}, \tilde{h})$ of an input. 
    In particular, we train a set-based classifier (e.g., DeepSets~\cite{zaheer2017deep}) based on $\mathcal{E}$. A set-based classifier takes a \emph{set} (i.e., $\mathcal{M}(\mathbf{x}, \tilde{h})$) as input and predicts a label  (1 or 0) for it. \SHF{A set-based classifier needs to be input-set-permutation-invariant, i.e., the predicted label does not rely on the order of the set elements. As a result, set-based classifiers and vector-based classifiers require substantially different neural network architectures. Moreover, set-based classification is generally harder than vector-based classification.} For simplicity, we use $f_s$ to denote the set-based classifier  and we use \emph{\SHF{EncoderMI}-S} to denote this \SHF{method}. 
    
     \item {\bf Threshold-based classifier (\SHF{EncoderMI}-T).} In a threshold-based classifier, we use the average similarity score in $\mathcal{M}(\mathbf{x}, \tilde{h})$ of an input to infer its membership. In particular, 
    our threshold-based classifier predicts an input to be a member if and only if the average similarity score in its membership features $\mathcal{M}(\mathbf{x}, \tilde{h})$ is no smaller than a threshold. The key challenge  is to determine the threshold with which the threshold-based classifier achieves high accuracy at membership inference.  
    Given a threshold $\theta$, we use $\alpha(\theta)$ (or $\beta(\theta)$) to denote the number of inputs in the shadow member (or non-member) dataset whose average similarity score in $\mathcal{M}(\mathbf{x}, \tilde{h})$ is smaller (or no smaller) than $\theta$. The accuracy of our threshold-based classifier with the threshold $\theta$ for the shadow dataset is $1 - (\alpha(\theta)+\beta(\theta))/|\mathcal{D}_s|$. Our threshold-based classifier uses the optimal threshold $\theta^*$ that maximizes such accuracy, i.e., minimizes $\alpha(\theta)+\beta(\theta)$. If we plot the probability distribution of the average similarity score for shadow members and shadow non-members as two curves, where the x-axis is average similarity score and y-axis is the probability that a random shadow member (or non-member) has the average similarity score, then the threshold $\theta^*$ is the intersection point of the two curves.

     Yeom et al.~\cite{yeom2018privacy} and Song et al.~\cite{song2019privacy} leveraged a similar threshold-based strategy for membership inference. Different from us, their \SHF{methods} were designed for classifiers and were based on the confidence scores outputted by a classifier. 
    
 \end{packeditemize}

\begin{algorithm}[!t]
    \caption{Our Membership Inference \SHF{Method EncoderMI} }
    \begin{algorithmic}[1]
    \REQUIRE $f_v$ (or $f_s$ or $f_t$), $h$, $\mathcal{A}$, $n$, $S$, and $\mathbf{x}$ \\
    \ENSURE \text{Member or non-member} \\
    \STATE $\{\mathbf{x}^1, \mathbf{x}^2, \cdots, \mathbf{x}^n\} \gets \textsc{Augmentation}(\mathbf{x},\mathcal{A},n)$ \\

    \STATE $\mathcal{M}(\mathbf{x},h) \gets \{S(h(\mathbf{x}^i), h(\mathbf{x}^j))| i \in [1, n], j \in [1,n], j > i\}$

    \IF {$f_v$}

    \RETURN $f_v(\textsc{Ranking}(\mathcal{M}(\mathbf{x},h)))$
    \ELSIF {$f_s$}
    \RETURN $f_s(\mathcal{M}(\mathbf{x},h))$
    \ELSIF {$f_t$}
    \RETURN $f_t(\textsc{Average}(\mathcal{M}(\mathbf{x},h)))$
    \ENDIF

\end{algorithmic}
\label{algorithml1}
\end{algorithm}

\subsection{Inferring Membership}
Given a black-box access to the target encoder $h$ and an input $\mathbf{x}$, we use the \SHF{inference} classifier $f_v$ (or $f_s$ or $f_t$) to predict whether the input $\mathbf{x}$ is a member of the target encoder $h$. Algorithm~\ref{algorithml1} shows our \SHF{method} . Given the input $\mathbf{x}$, the data augmentation module $\mathcal{A}$, and an integer $n$, the function \textsc{Augmentation} produces $n$ augmented inputs. We use the target encoder $h$ to produce a feature vector for each augmented input, and then we compute the set of pairwise similarity scores as the membership features $\mathcal{M}(\mathbf{x}, h)$ for the input $\mathbf{x}$. Finally, we use the \SHF{inference} classifier to  infer the membership status of the input $\mathbf{x}$ based on the extracted membership features. The function \textsc{Ranking} ranks the similarity scores in $\mathcal{M}(\mathbf{x}, h)$ in a descending order, while the function \textsc{Average} computes the average of the similarity scores in $\mathcal{M}(\mathbf{x}, h)$. 

\section{Evaluation}
\label{evaluation-section}
We evaluate \SHF{EncoderMI} on image encoders pre-trained on unlabeled images in this section. In Section~\ref{attack_clip_section}, we apply \SHF{EncoderMI} to  CLIP, which was pre-trained on unlabeled (image, text) pairs. 
\subsection{Experimental Setup}
\myparatight{Datasets}
We conduct our experiments on CIFAR10, STL10, and Tiny-ImageNet datasets. 

\begin{packeditemize}
    \item {\bf CIFAR10~\cite{krizhevsky2009learning}.} CIFAR10 dataset contains 60,000 colour images from 10 object categories. In particular, the dataset contains 50,000 training images and 10,000 testing images. The size of each image is $32 \times 32$.
    
    \item {\bf STL10~\cite{coates2011analysis}.} STL10 dataset contains 13,000 labeled colour images from 10 classes. Specifically, the dataset is divided into 5,000 training images and 8,000 testing images. We note that STL10 dataset also contains 100,000 unlabeled images. The size of each image is $96 \times 96$ in this dataset.  
    
    \item {\bf Tiny-ImageNet~\cite{tinyimagenet}.} 
    Tiny-ImageNet dataset contains 100,000 training images and 10,000 testing images from 200 classes.  Each class has 500 training images and 50 testing images.  Each image has size $64 \times 64$.
\end{packeditemize}

\myparatight{Training target encoders} For CIFAR10 or Tiny-ImageNet, we randomly sample 10,000  images from its training  data  as the pre-training dataset to train a target encoder; and for STL10, we randomly sample 10,000  images from its  unlabeled  data  as the pre-training dataset.  By default, we use ResNet18~\cite{he2016deep} as the  architecture for the target encoder. Moreover, we use MoCo~\cite{he2020momentum} to pre-train the target encoder on a pre-training dataset.  We adopt the publicly available implementation of MoCo v1~\cite{mococode} with  the default parameter setting when pre-training our target encoders. Unless otherwise mentioned, we train a target encoder for 1,600 epochs. 
For CIFAR10 or Tiny-ImageNet, we treat its 10,000 testing images as ground truth ``non-member" of the target encoder. For STL-10, we treat its 5,000 training images and the first 5,000 testing images  as ``non-member" of the target encoder. Therefore, unless otherwise mentioned, for each target encoder, we have 10,000 ground truth members and 10,000 ground truth non-members.

\myparatight{Training shadow encoders} In the scenario where the \SHF{inferrer} knows the pre-training data distribution of the target encoder, we randomly sample 20,000 images from the training or unlabeled data of the corresponding dataset as the shadow dataset. In the scenario where the \SHF{inferrer} does not know the pre-training data distribution, we randomly sample 20,000 images from the training data of CIFAR10 as the shadow dataset when the pre-training dataset is from STL-10, and we randomly sample 20,000 images from the unlabeled data of STL-10 as the shadow dataset when the pre-training dataset is CIFAR10  or Tiny-ImageNet. 

We randomly split a shadow dataset into two disjoint sets, i.e., \emph{shadow member dataset} and \emph{shadow non-member dataset}, each of which contains 10,000 images. We train a shadow encoder using a shadow member dataset. 
We adopt the same architecture (i.e., ResNet18) for a shadow encoder if the \SHF{inferrer} knows the  architecture of the target encoder and use VGG-11~\cite{simonyan2014very} with batch normalization otherwise. We adopt the same training algorithm (i.e., MoCo) to pre-train a shadow encoder if the \SHF{inferrer} knows the  algorithm used to pre-train the target encoder and adopt SimCLR~\cite{chen2020simple} otherwise. We use the publicly available implementations~\cite{mococode, simclr-code} with the default parameter settings for both training algorithms in our experiments. We train each shadow encoder for 1,600 epochs.

\myparatight{Building \SHF{inference} classifiers} We build \SHF{inference} classifiers based on a shadow dataset and a shadow encoder.   \SHF{EncoderMI}-V uses a vector-based \SHF{inference} classifier. We use a fully connected neural network with two hidden layers as our vector-based classifier. In particular, the number of neurons in both hidden layers are  256.  \SHF{EncoderMI}-S uses a set-based \SHF{inference} classifier. We choose DeepSets~\cite{zaheer2017deep} as our set-based \SHF{inference} classifier. Moreover, we adopt the publicly available code for DeepSets~\cite{deepsets} in our implementation. For both the vector-based classifier and the set-based classifier, we adopt cross-entropy as loss function and use Adam optimizer with initial learning rate of 0.0001 to train for 300 epochs. Note that  \SHF{EncoderMI}-T leverages a  threshold-based classifier and does not require training.

\myparatight{Evaluation metrics}
Following previous work~\cite{shokri2017membership,salem2018ml}, we adopt \emph{accuracy}, \emph{precision}, and \emph{recall} to evaluate membership inference methods. Given an evaluation dataset that contains ground truth members and non-members of the target encoder,  accuracy of a \SHF{method} is the ratio of the ground truth members/non-members that are correctly predicted by the \SHF{method}; precision of a \SHF{method} is the fraction of its predicted members that are indeed members; and recall of a \SHF{method} is the fraction of ground truth members that are predicted as members by the method.

\myparatight{Compared methods} Existing membership inference \SHF{methods~\cite{shokri2017membership,salem2018ml,song2019privacy,song2020information,choo2020label}} aim to infer members of a classifier \SHF{or a text embedding model}. We generalize these \SHF{methods} to the \SHF{contrastive learning} setting as baseline \SHF{methods}. In particular, 
we compare our \SHF{methods} with the following \SHF{five} baseline \SHF{methods, where the first three are for downstream classifiers, while the last two are for encoders.} 

    {\bf Baseline-A.} The target encoder is used to train a downstream classifier for a downstream task. Therefore, \SHF{in this baseline  method}, we use the target encoder to train a downstream classifier (called \emph{target downstream classifier}) for a downstream task, and then we apply existing membership inference \SHF{methods}~\cite{shokri2017membership,salem2018ml} to the target downstream classifier. In particular, we consider 
    CIFAR10 as a downstream task and we randomly sample 10,000 of its training examples as the downstream dataset. The downstream dataset does not have overlap with the pre-training dataset and the shadow dataset. Given a shadow encoder and the downstream dataset, we train a downstream classifier (called \emph{shadow downstream classifier}) via using the shadow encoder as a feature extractor. 
    We query the confidence score vector for each input in the shadow member (or non-member) dataset outputted by the shadow downstream classifier and label it as ``member" (or ``non-member"). Given these confidence score vectors as well as the corresponding labels, we train a vector-based \SHF{inference} classifier. 
    For a given input, we first query its confidence score vector outputted by the target downstream classifier and then use the \SHF{inference} classifier to predict whether it's a member of the target encoder. Note that, following previous work~\cite{salem2018ml}, we rank the confidence scores for an input, which outperforms unranked confidence scores.

    {\bf \SHF{Baseline-B}.} \SHF{Choquette-Choo et al.~\cite{choo2020label} proposed label-only membership inference to a classifier. 
    Roughly speaking, they construct a binary feature vector for an input based on some augmented versions of the input.  An entry of the feature vector is 1 if and only if the corresponding augmented version is predicted correctly by the target classifier.  
     This label-only membership inference method requires the ground truth label of an input. The pre-training data are unlabeled in contrastive learning, making the method not applicable to infer members of an encoder in practice. However, since the pre-training data CIFAR10 and Tiny-ImageNet have ground truth labels in our experiments, we assume an inferrer knows them and we evaluate the label-only method. Note that we cannot evaluate this method when the pre-training dataset is from  STL10 as they are unlabeled. 
    Similar to Baseline-A, we also apply this method to a target downstream classifier to infer members of the target encoder. For each input $\mathbf{x}$ in the shadow dataset, we create $e$ augmented inputs. Moreover, we 
    use the shadow downstream classifier to predict the label of $\mathbf{x}$ and each augmented input.  
    We construct a binary vector $(b_0, b_1, b_2,\cdots,b_e)$ as the membership features for $\mathbf{x}$, where $b_0=1$ (or $b_i=1$) if and only if the shadow downstream classifier correctly predicts the label of  $\mathbf{x}$ (or the $i$th augmented input), where $i=1, 2, \cdots, e$.  We label the membership features of an input as ``member" (or ``non-member") if the input is in the shadow member (or non-member) dataset. Given membership features and their labels, we train a vector-based inference classifier. Then, we use the inference classifier to infer members of a target encoder via a target downstream classifier. We set $e=10$ in our experiments. 
    }

    {\bf \SHF{Baseline-C.}} \SHF{Song et al.~\cite{song2019privacy} developed adversarial example based membership inference methods against classifiers. Specifically, they leverage the confidence scores produced by the target classifier for adversarial examples crafted from an input to infer whether the input is in the training dataset of the target classifier. For instance, their targeted adversarial example based method (discussed in Section 3.3.1 of~\cite{song2019privacy}) first crafts $k-1$ targeted adversarial examples for an input (one targeted adversarial example for each label that is not the input's ground truth label), then uses the target classifier to compute confidence scores for each of them, and finally concatenates the confidence scores as membership features for the input, where $k$ is the number of classes in the target classifier. They train an inference classifier for each class of the target classifier and use the inference classifier corresponding to the ground truth label of an input to infer its membership. 
    
    Adversarial example of an input can be viewed as the input's augmented version. Therefore, we consider these methods in our experiments. We note that  these methods  require the ground truth label of an input and the ground truth label is one class of the downstream classifier. However, in contrastive learning, pre-training data often do not have labels. Moreover, even if the pre-training data  have labels, their labels may not be the same as those of the downstream classifier. Therefore, we adapt the targeted adversarial example based method  to a downstream classifier in our setting. Specifically, given an input $\mathbf{x}$, we use PGD~\cite{madry2018towards} to generate $k$ targeted adversarial examples based on a shadow downstream classifier, where $k$ is the number of classes of the shadow downstream classifier. We then obtain $k$ confidence score vectors outputted by the shadow downstream classifier for the $k$ targeted adversarial examples, and we concatenate them as membership features for $\mathbf{x}$. Finally, we train one vector-based inference classifier based on the membership features of inputs in the shadow dataset, and we apply it to infer members of the target encoder via the target downstream classifier. Moreover, following~\cite{song2019privacy}, we set the perturbation budget (i.e., $\epsilon$) to be $8/255$ when generating targeted adversarial examples. 
  }
    
    \begin{table}[!t]\renewcommand{\arraystretch}{1}
     \centering
     \caption{Accuracy, precision, and recall (\%) of \SHF{the five baseline methods. ``--'' means not applicable.}}
     \setlength{\tabcolsep}{1mm}

\vspace{-2mm}
\subfloat[Baseline-A]{
          \begin{tabular}{|c|c|c|c|}
     \hline
     Pre-training dataset & Accuracy  &Precision   &Recall  \\
     \hline
     CIFAR10  &55.1 &53.4 &73.1  \\ \hline
     STL10 &54.3 &53.7 &62.2  \\  \hline
     Tiny-ImageNet  &47.3 &48.2 &68.3\\  
     \hline
     \end{tabular}
      \label{tab: baseline-D}
     } \\ 
     \vspace{-2mm}
     \SHF{
\subfloat[Baseline-B]{
          \begin{tabular}{|c|c|c|c|}
     \hline
     Pre-training dataset & Accuracy  &Precision   &Recall  \\
     \hline
     CIFAR10  &54.6 &63.1 &58.2  \\ \hline
     STL10 &-- &-- &--  \\  \hline
     Tiny-ImageNet  &51.8 &53.7 &47.6\\  
     \hline
     \end{tabular}
      \label{tab: baseline-D} 
     } \\ 
     \vspace{-2mm}
\subfloat[Baseline-C]{
          \begin{tabular}{|c|c|c|c|}
     \hline
     Pre-training dataset & Accuracy  &Precision   &Recall  \\
     \hline
     CIFAR10  &52.8 &54.1 &43.1  \\ \hline
     STL10 &50.5 &50.1 &57.9  \\  \hline
     Tiny-ImageNet  &50.2 &52.1 &42.3\\  
     \hline
     \end{tabular}
      \label{tab: baseline-D}
     }\vspace{-2mm}
     }          
          \subfloat[Baseline-D]{
          \begin{tabular}{|c|c|c|c|}
     \hline
     Pre-training dataset & Accuracy  &Precision   &Recall  \\
     \hline
     CIFAR10  &50.7 &50.6  &51.0  \\ \hline
     STL10  &50.1 &49.9   &50.3  \\  \hline
     Tiny-ImageNet  &49.5  &49.3   &49.2 \\  
     \hline
     \end{tabular}
      \label{tab: baseline-E}
     }
     \\
          \vspace{-2mm}
          \SHF{
\subfloat[Baseline-E]{
          \begin{tabular}{|c|c|c|c|}
     \hline
     Pre-training dataset & Accuracy  &Precision   &Recall  \\
     \hline
     CIFAR10  &64.5 &63.8 &67.2  \\ \hline
     STL10 &67.0 &65.7 &71.3  \\  \hline
     Tiny-ImageNet  &68.6 &67.8 &70.8\\  
     \hline
     \end{tabular}
      \label{tab: baseline-D}
     } \\
     }

     \label{tab: baseline-attack}
     \vspace{-7mm}
 \end{table}

        {\bf Baseline-D.} In this baseline \SHF{method}, we consider that an \SHF{inferrer} treats the target encoder as if it was a classifier. In other words, the \SHF{inferrer} treats the feature vector outputted by the target encoder for an input as if it was a \emph{confidence score vector} outputted by a classifier. Therefore, we can apply the confidence score vector based \SHF{methods}~\cite{shokri2017membership,salem2018ml} to \SHF{infer  members of} the target encoder. Specifically, 
    given a  shadow encoder, we use it to output a feature vector for each input in the corresponding shadow dataset. Moreover, we assign label ``member" (or ``non-member") to the feature vector of an input in the shadow member (or non-member) dataset. Then, we train a vector-based \SHF{inference} classifier using the feature vectors and their labels.  
    Given a target encoder and an input, we first obtain the feature vector produced by the target encoder for the input. Then, based on the feature vector,  the \SHF{inference} classifier predicts the input to be a member or non-member of the target encoder.

    {\bf \SHF{Baseline-E.}} \SHF{Song et al.~\cite{song2020information} studied membership inference to embedding models in the text domain. Their method can be used to infer whether a sentence is in the training dataset of a text embedding model.  In particular, they use the average cosine similarity between the embedding vectors of the center word and each remaining word in a sentence to infer the membership of the sentence. We extend this method to our setting. Specifically, we could view an image as a "sentence" and each patch of an image as a "word". Then, we can use an image encoder to produce a feature vector for each patch. We view the center patch as the center word and compute its cosine similarity with each remaining patch. Finally, we use the average similarity score to infer the membership of the original image. Specifically, the image is predicted as a member  if the average similarity score is larger than a threshold. Similar to our EncoderMI-T, we use a shadow dataset to determine the optimal threshold, i.e., we use the threshold that achieves the largest inference accuracy on the shadow dataset. We evenly divide an image into $3\times 1$ (i.e., 3), $3\times 3$ (i.e., 9), or $3\times 5$ (i.e., 15) disjoint patches in our experiments. We found $3\times 3$ achieves the best performance, so we will show results for $3\times 3$ in the main text and defer the results for $3\times 1$ and $3\times 5$ to Appendix. 
    }

\begin{figure}[!tb]
\centering
 \vspace{-3mm}
\subfloat[]{\includegraphics[width=0.25\textwidth]{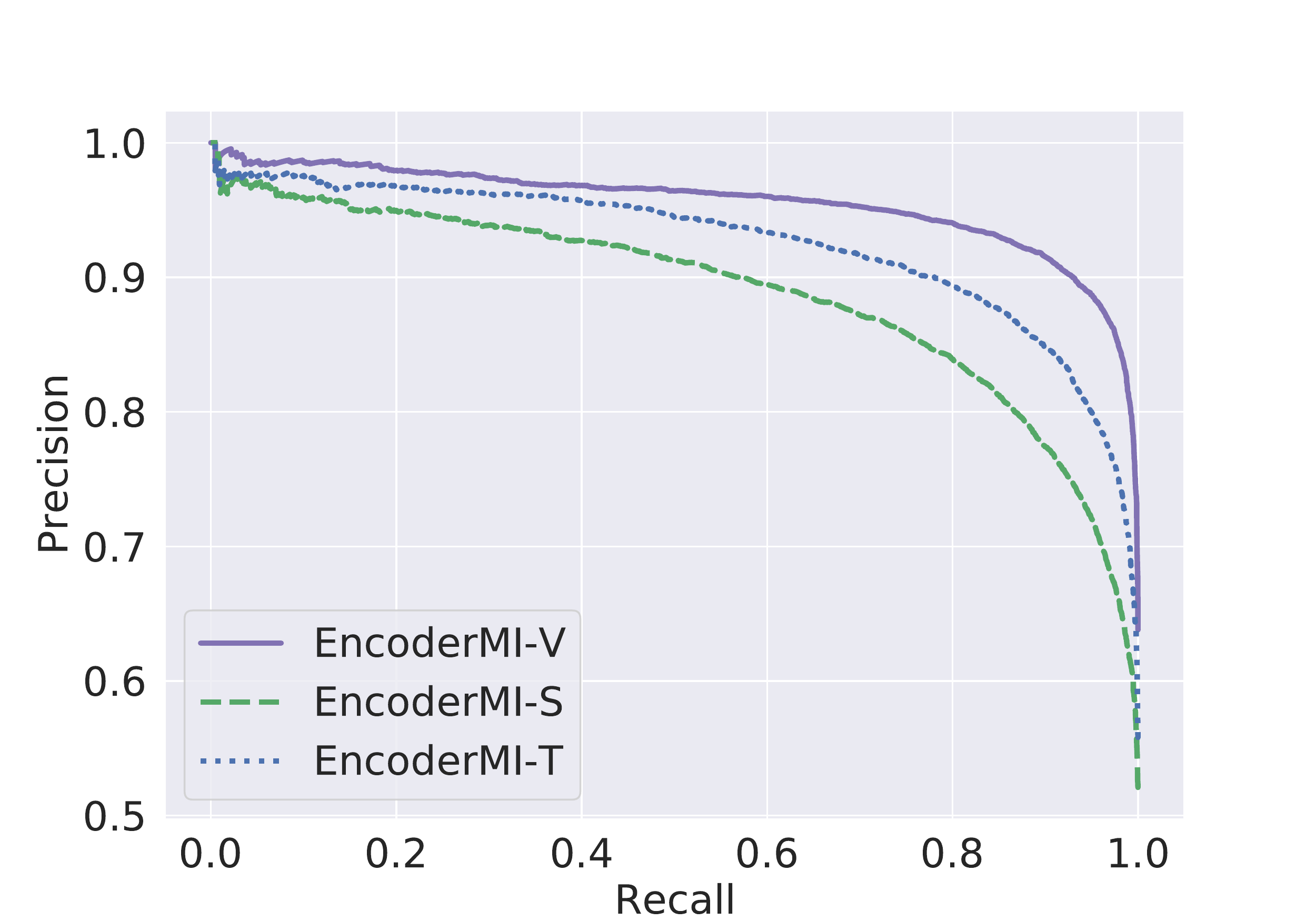}\label{prcurve}}
\subfloat[]{\includegraphics[width=0.25\textwidth]{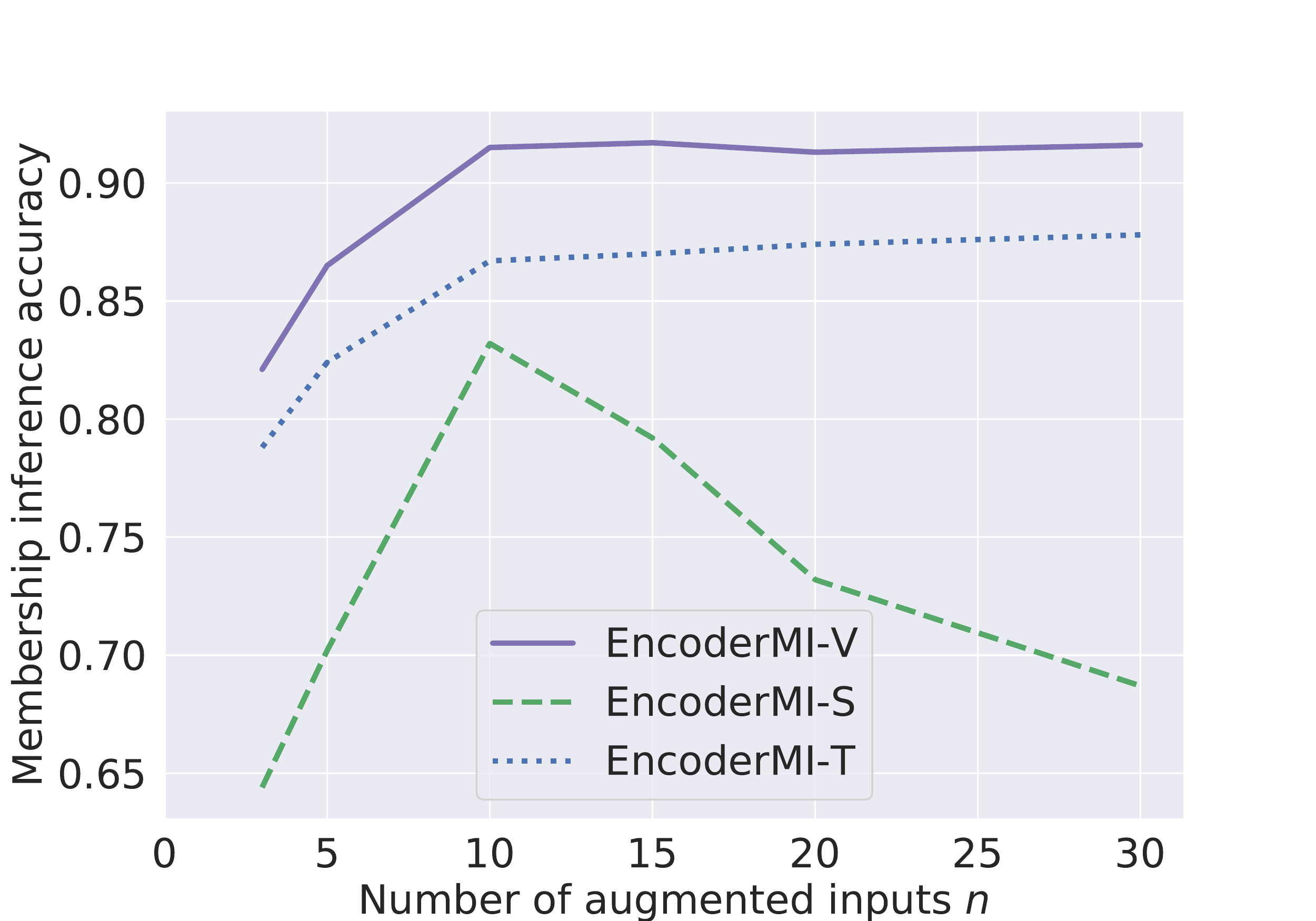}\label{impact-n}}

 \vspace{-3mm}
\caption{\SHF{(a) Precision-recall trade-off.} (b) Impact of $n$ on accuracy. The dataset is CIFAR10. }
 \vspace{-4mm}
\end{figure}

\myparatight{Parameter settings} We adopt the following default parameters for our \SHF{method}: 
we set $n=10$ and we adopt cosine similarity as our similarity metric $S$ since all contrastive learning algorithms use cosine similarity to measure similarity between two feature vectors. By default, we assume the \SHF{inferrer} knows the pre-training data distribution, the encoder architecture, and the training algorithm of the target encoder. Unless otherwise mentioned, we show results on CIFAR10 as the pre-training dataset.  When the inferrer does not know the target encoder's training algorithm, we assume the inferrer uses random resized crop only to obtain augmented versions of an input when querying the target encoder because we found such data augmentation achieves the best performance. Note that we resize each image in STL10 and Tiny-ImageNet to $32 \times 32$ to be consistent with CIFAR10.

\subsection{Experimental Results}
\label{exp-results}

\begin{table*}[h]\renewcommand{\arraystretch}{1}
      \centering
      \small
      \setlength{\tabcolsep}{1mm}
      \caption{Average accuracy, precision, and recall (\%) of our \SHF{methods} for the target encoder pre-trained on CIFAR10 dataset. $\surd$ (or $ \times$) means the \SHF{inferrer} has
  (or does not have) access to the corresponding background knowledge of the target encoder. The numbers in parenthesis are standard deviations in 5 trials.  }
  \vspace{-2mm}

     \begin{tabular}{|c|c|c|c|c|c|c|c|c|c|c|c|}
      \hline
     \multirow{3}{*}{\makecell{Pre-training\\data distribution}} & \multirow{3}{*}{\makecell{Encoder\\architecture}} & \multirow{3}{*}{\makecell{Training\\ algorithm}} &\multicolumn{3}{c|}{Accuracy} &\multicolumn{3}{c|}{Precision} &\multicolumn{3}{c|}{Recall} \cr\cline{4-12}  
      & & & \makecell{ \SHF{Encod-}\\ \SHF{erMI}-V} & \makecell{ \SHF{Encod-}\\ \SHF{erMI}-S} & \makecell{ \SHF{Encod-}\\ \SHF{erMI}-T} &  \makecell{ \SHF{Encod-}\\ \SHF{erMI}-V} & \makecell{ \SHF{Encod-}\\ \SHF{erMI}-S} & \makecell{ \SHF{Encod-}\\ \SHF{erMI}-T} 	&  \makecell{ \SHF{Encod-}\\ \SHF{erMI}-V} &\makecell{ \SHF{Encod-}\\ \SHF{erMI}-S} & \makecell{ \SHF{Encod-}\\ \SHF{erMI}-T}\\ \hline
          $\times$ & $\times$ & $\times$ 	& 86.2 (2.04)  & 78.1 (2.21) 	& 82.1 (1.91) 	& 87.8 (1.15)   & 78.9 (1.69) 	& 80.1 (1.01) 	& 89.3 (2.15)   & 86.8 (2.44)  	& 87.2 (1.89) \\ \hline
          $\surd$  & $\times$ & $\times$ 	& 86.9 (2.03)  & 79.6 (2.05) 	& 83.3 (1.75)	& 88.3 (1.64)   & 79.8 (1.34)   & 81.0 (1.12) 	& 90.9 (2.44)   & 87.4 (2.51)   & 89.2 (1.63)\\ \hline
          $\times$ & $\surd$  & $\times$ 	& 87.0 (1.21)  & 79.4 (1.39)	& 83.5 (1.03) 	& 88.4 (1.37)   & 79.6 (0.98) 	& 81.6 (0.87)	& 91.4 (1.17)  & 87.2 (1.46)   	& 87.9 (0.92)\\ \hline
          $\times$ & $\times$ & $\surd$  	& 86.7 (0.81)  & 79.2 (1.05)	& 83.0 (0.77) 	& 88.2 (0.83)   & 79.9 (1.10) 	& 80.4 (0.81)	& 91.4 (0.76)  & 87.1 (1.01) 	& 89.1 (0.79)\\ \hline
          $\surd$  & $\surd$  & $\times$ 	& 87.2 (2.17)  & 79.9 (1.88)    & 83.6 (1.66)  	& 88.4 (1.38)   & 80.1 (1.03)   & 81.9 (0.98) 	& 91.5 (1.23)	& 87.9 (1.32)  	& 87.9 (1.01) \\  \hline
          $\surd$  & $\times$ & $\surd$  	& 87.6 (0.45)  & 80.3 (0.47) 	& 84.2 (0.39) 	& 88.7 (0.43)   & 80.6 (0.44)   & 83.1 (0.37) 	& 91.7 (0.51)   & 87.7 (0.49)  	& 88.0 (0.46) \\ \hline
          $\times$ & $\surd$  & $\surd$  	& 90.2 (0.37)  & 81.1 (0.43)  	& 85.2 (0.37) 	& 89.5 (0.31)   & 80.5 (0.39) 	& 85.1 (0.33)	& 93.3 (0.32)   & 88.2 (0.48)  	& 88.9 (0.38)\\ \hline
          $\surd$  & $\surd$  & $\surd$  	& 91.4 (0.28)  & 83.1 (0.27)	& 86.6 (0.28) 	& 90.1 (0.23)   & 80.8 (0.29) 	& 85.9 (0.27) 	& 93.5 (0.22)   & 88.3 (0.30) 	& 89.1 (0.25) \\ \hline
      \end{tabular}

      \label{tab: attack-results-cifar}
   \vspace{-2mm}   
  \end{table*}

\myparatight{Existing membership inference \SHF{methods} are insufficient}
\SHF{Table~\ref{tab: baseline-attack} shows the accuracy, precision, and recall of the five baseline methods.} Note that we consider an \SHF{inferrer} with the strongest background knowledge in our threat model, i.e., the \SHF{inferrer} knows the pre-training data distribution, the encoder architecture, and the training algorithm. In other words, the shadow encoders are trained in the background knowledge  $\mathcal{B}=(\surd,\surd,\surd)$. 
We find that the accuracies of  Baseline-A,  Baseline-B, Baseline-C, and Baseline-D  are close to 50\%, i.e., their accuracies are close to that of random guessing  in which an input is  predicted as a member or non-member with probability 0.5. The reason is that they were designed to infer members of a classifier instead of an encoder.  The confidence score vector can capture whether the classifier is overfitted for the input, while the feature vector itself does not capture whether the encoder is overfitted for the input. As a result, these membership inference \SHF{methods} can infer the members of a classifier but not an encoder. \SHF{Baseline-E is better than random guessing. The reason is that a patch of an input can be viewed as an augmented version of the input, and the similarity scores between patches capture the overfitting of an image encoder to some extent. However, the accuracy of Baseline-E is still low, compared to our EncoderMI.}

\myparatight{Our \SHF{methods} are effective} Table~\ref{tab: attack-results-cifar},~\ref{tab: attack-results-stl} (in Appendix), and~\ref{tab: attack-results-tinyimagenet} (in Appendix) show the accuracy, precision, and recall of our \SHF{methods} under the 8 different types of background knowledge for CIFAR10, STL-10, and Tiny-ImageNet datasets, respectively. \SHF{The results are averaged in five trials.} 
First, our \SHF{methods} are effective under all the 8 different types of background knowledge.   
For instance, our \SHF{EncoderMI}-V can achieve 88.7\% -- 96.5\% accuracy on Tiny-ImageNet  under the 8 types of background knowledge. Second, we find that  \SHF{EncoderMI}-V is more effective than \SHF{EncoderMI}-S and \SHF{EncoderMI}-T in most cases. In particular, \SHF{EncoderMI}-V achieves higher accuracy (or precision or recall) than \SHF{EncoderMI}-S and \SHF{EncoderMI}-T in most cases. We suspect that \SHF{EncoderMI}-V outperforms \SHF{EncoderMI}-S because set-based classification is generally more challenging than vector-based classification, and thus viewing  membership inference as a vector-based classification problem can achieve better inference performance. Our \SHF{EncoderMI}-T can achieve similar accuracy, precision, and recall with \SHF{EncoderMI}-S, which means that the average pairwise cosine similarity score for an input image already provides rich  information about the input's membership status.  Third, our \SHF{methods} achieve  higher recall than precision, i.e., our \SHF{methods} predict more inputs as members than non-members. \SHF{Fourth, the standard deviations tend to be larger when the inferrer has less background knowledge. This is because membership inference is less stable with less background knowledge. }

\SHF{Figure~\ref{prcurve} shows the precision-recall trade-off of our three methods under the background knowledge $\mathcal{B}=(\surd,\surd,\surd)$. The curves are obtained via tuning the classification thresholds in the three inference classifiers to produce different precisions and recalls. Our results show that precision drops slightly as recall increases up to around 0.9, and then drops sharply as recall further increases. }

\myparatight{Impact of the \SHF{inferrer}'s background knowledge} Based on Table~\ref{tab: attack-results-cifar},~\ref{tab: attack-results-stl}, and~\ref{tab: attack-results-tinyimagenet}, we have three major observations about the impact of the \SHF{inferrer}'s background knowledge on our \SHF{methods}. 
First,  \SHF{EncoderMI}-V achieves higher accuracy as the \SHF{inferrer} has access to more background knowledge, and we have the same observation for \SHF{EncoderMI}-S and \SHF{EncoderMI}-T in most cases. For instance,  \SHF{EncoderMI}-V achieves 96.5\% accuracy when the \SHF{inferrer} knows all the three dimensions of background knowledge while achieving 88.7\% accuracy when the \SHF{inferrer} does not know any of them for Tiny-ImageNet dataset. Second, among the three dimensions of background knowledge, training algorithm is the most informative for STL10 and Tiny-ImageNet, while the three dimensions contribute equally for CIFAR10. For instance, on Tiny-ImageNet,  \SHF{EncoderMI}-V achieves 94.1\% accuracy when the \SHF{inferrer} only has access to the training algorithm, while  \SHF{EncoderMI}-V respectively achieves 89.1\% and 93.0\% accuracy when the \SHF{inferrer} only has access to the encoder architecture and the pre-training data distribution. For CIFAR10, \SHF{EncoderMI}-V achieves around 86.8\% accuracy when the \SHF{inferrer} only has access to any of the three dimensions. Third, there is no clear winner among the encoder architecture and pre-training data distribution. For instance, having access to the pre-training data distribution alone (i.e., $\mathcal{B}=(\surd,\times,\times)$) achieves higher accuracy than having access to the encoder architecture alone (i.e., $\mathcal{B}=(\times,\surd,\times)$) for all our three \SHF{methods} on Tiny-ImageNet, but we observe the opposite for \SHF{EncoderMI}-V and \SHF{EncoderMI}-T on STL10.

 \begin{figure*}[!t]
\centering
\vspace{-3mm}
\subfloat[\SHF{EncoderMI}-V]{\includegraphics[width=0.33\textwidth]{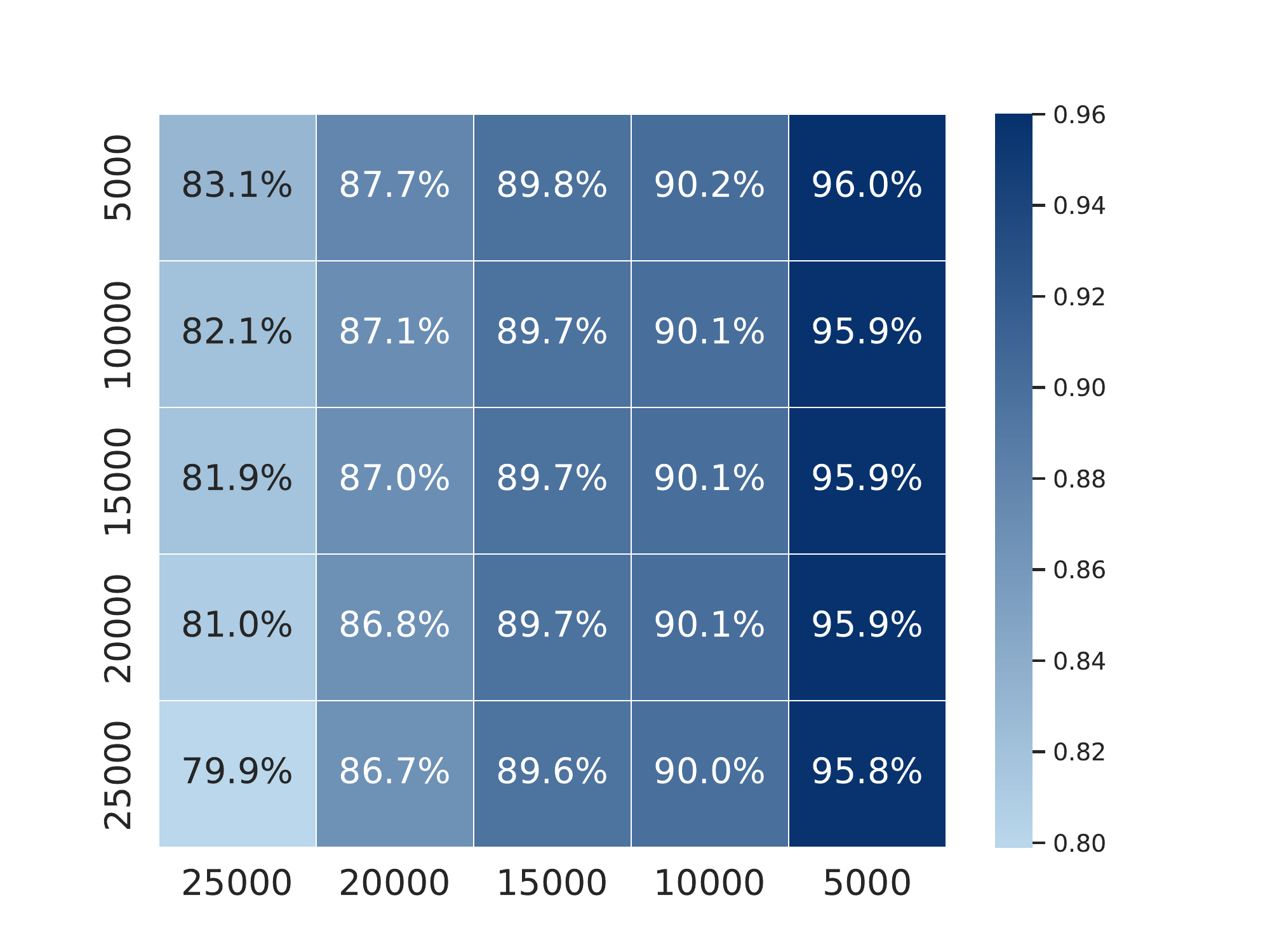}}
\subfloat[\SHF{EncoderMI}-S]{\includegraphics[width=0.33\textwidth]{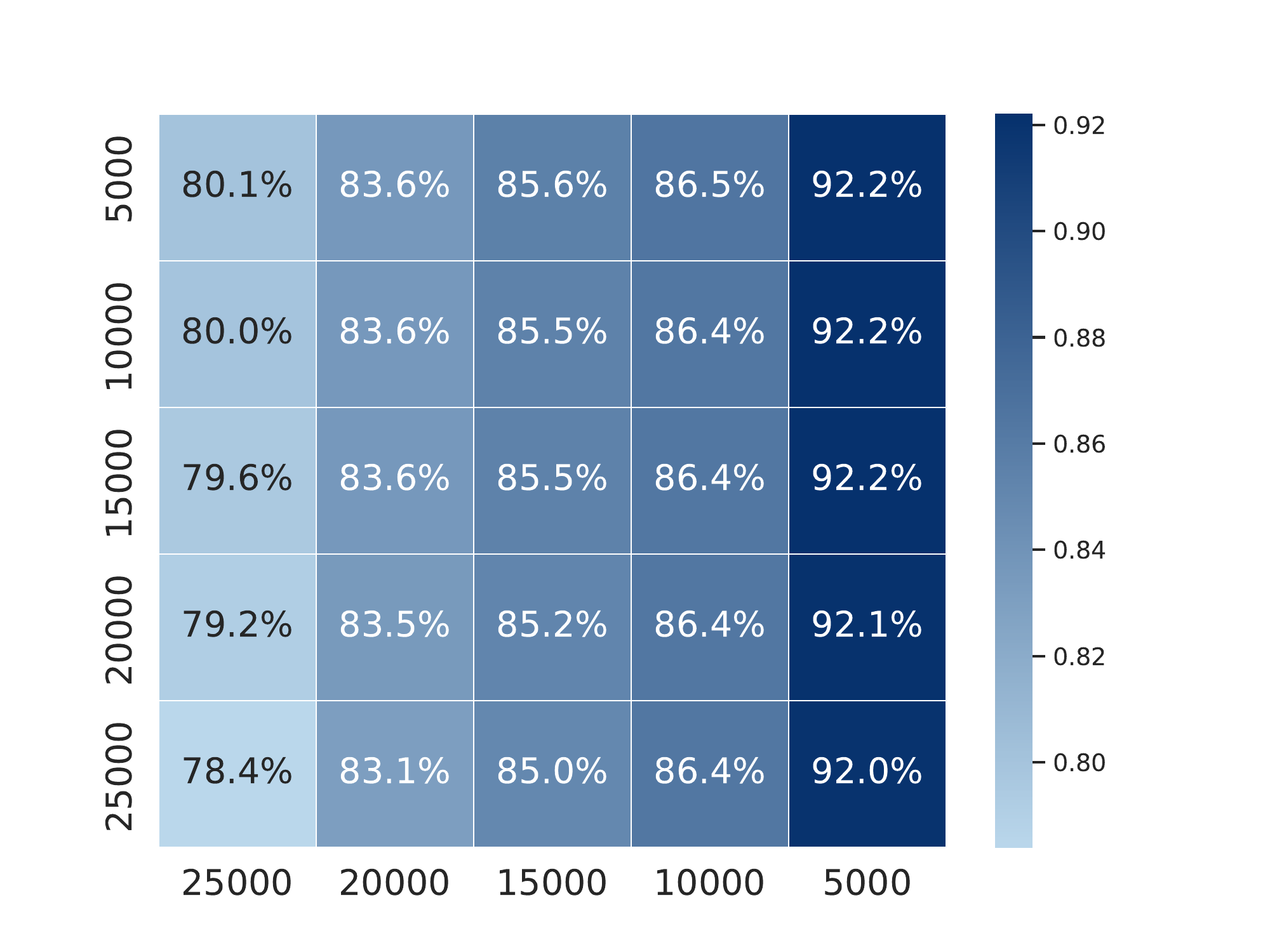}}
\subfloat[\SHF{EncoderMI}-T]{\includegraphics[width=0.33\textwidth]{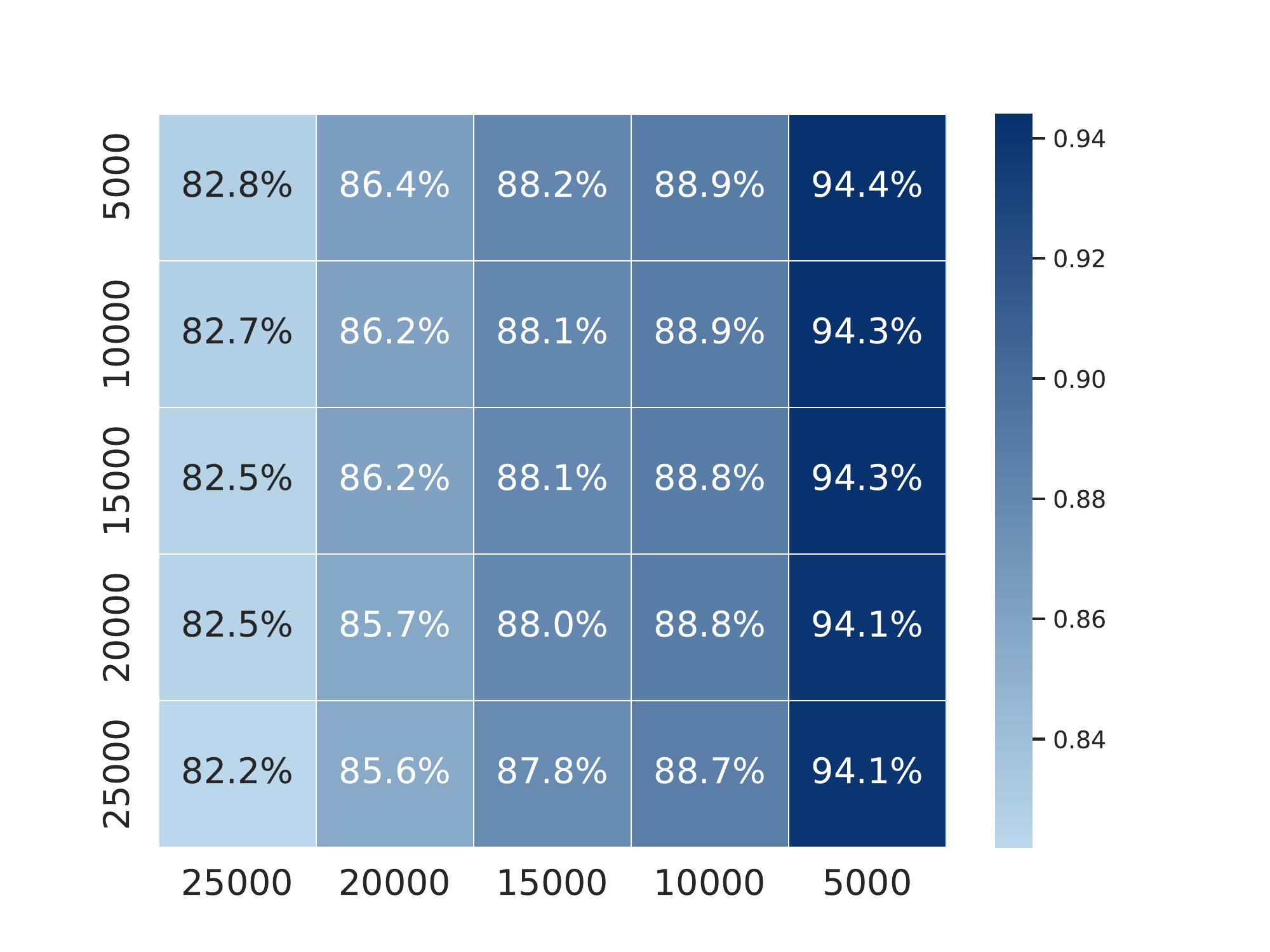}}
\vspace{-3mm}
\caption{Impact of the size of the pre-training dataset (x-axis) and the shadow dataset (y-axis) on the accuracy of membership inference. Both the pre-training dataset and shadow dataset are randomly sampled from STL10.}
\vspace{-3mm}
\label{fig:vary-size}
\end{figure*}

 \begin{table}[!t]\renewcommand{\arraystretch}{1}
    \small
     \centering
          \caption{Accuracy, precision, and recall (\%) of our \SHF{methods} with different similarity metrics. The dataset is CIFAR10. }
     \begin{tabular}{|c |c| c| c|c|}
     \hline
    \makecell{\SHF{Method}} & \makecell{Similarity \\metric}  & Accuracy &Precision &Recall\\
     \hline
     \multirow{3}{*}{\SHF{EncoderMI}-V} & \makecell{Cosine} &91.5 &90.0 &93.5 \cr\cline{2-5} 
     & \makecell{Correlation} &89.3 &87.9 &92.2 \cr\cline{2-5} 
     & \makecell{Euclidean }  & 88.9& 85.3 &  94.5\cr\cline{1-5}  
     \multirow{3}{*}{\SHF{EncoderMI}-S} & \makecell{Cosine} &83.2 &80.5 &87.9 \cr\cline{2-5} 
     & \makecell{Correlation} &76.3 &75.5 &78.5 \cr\cline{2-5} 
     & \makecell{Euclidean }  & 75.8& 73.6 &  81.4\cr\cline{1-5}       
     \multirow{3}{*}{\SHF{EncoderMI}-T} & \makecell{Cosine} &86.7 &85.7 &89.0 \cr\cline{2-5} 
     & \makecell{Correlation} &80.6 &79.8 &81.6 \cr\cline{2-5} 
     & \makecell{Euclidean}  & 80.7& 80.1 &  82.5\cr\cline{1-5}  
     \end{tabular}
\vspace{-4mm}
     \label{tab: diff-similarity}
 \end{table}

\myparatight{Impact of $n$}
Figure~\ref{impact-n} shows the impact of the number of augmented inputs $n$ on the accuracy of our \SHF{methods} for CIFAR10, where the \SHF{inferrer}'s background knowledge is $\mathcal{B}=(\surd,\surd,\surd)$. 
We observe that, for \SHF{EncoderMI}-V and \SHF{EncoderMI}-T, the accuracy first increases and then saturates as $n$ increases. However, for \SHF{EncoderMI}-S, the accuracy first increases and then decreases as $n$ increases.    
We suspect the reason is that as $n$ increases, the number of pairwise similarity scores in the membership features increases exponentially, making set-based classification harder.

\myparatight{Impact of the similarity metric $S$} Table~\ref{tab: diff-similarity} shows the impact of the similarity metric $S$ on our \SHF{methods}, where ``Correlation'' refers to Pearson correlation coefficient. We have two observations from the experimental results. First, the cosine similarity metric achieves the highest accuracy (or precision or recall). The reason is that the cosine similarity metric is also used in the pre-training of the target encoder. Second, our \SHF{methods} still achieve high accuracy (or precision or recall) when using different similarity metrics from the one used in the pre-training of the target encoder.

\myparatight{Impact of the size of the pre-training and shadow datasets} Figure~\ref{fig:vary-size} shows the impact of the size of the pre-training dataset and the shadow dataset on the accuracy of our three \SHF{methods}, where  the \SHF{inferrer}'s background knowledge is $\mathcal{B}=(\surd,\surd,\surd)$. 
Both the pre-training dataset and shadow dataset are randomly sampled from the unlabeled data of STL10, but they do not have overlaps. Note that we did not use CIFAR10 in these experiments  because its dataset size is small and we cannot sample disjoint  pre-training and shadow datasets with large sizes. First, we observe that the accuracy of our \SHF{methods} increases as the pre-training dataset becomes smaller. The reason is that the target encoder is more overfitted to the pre-training dataset when its size is smaller. 
Second, our \SHF{methods} are less sensitive to the shadow dataset size. In particular, given a pre-training dataset size, each of our three \SHF{methods} achieves similar accuracy when the shadow dataset size ranges between 5,000 and 25,000. Our results show that the shadow encoder pre-trained on shadow dataset with various sizes can mimic the  behavior of the target encoder in terms of membership inference.

\begin{figure}[!tb]
\centering
\vspace{-2mm}
\includegraphics[width=0.3\textwidth]{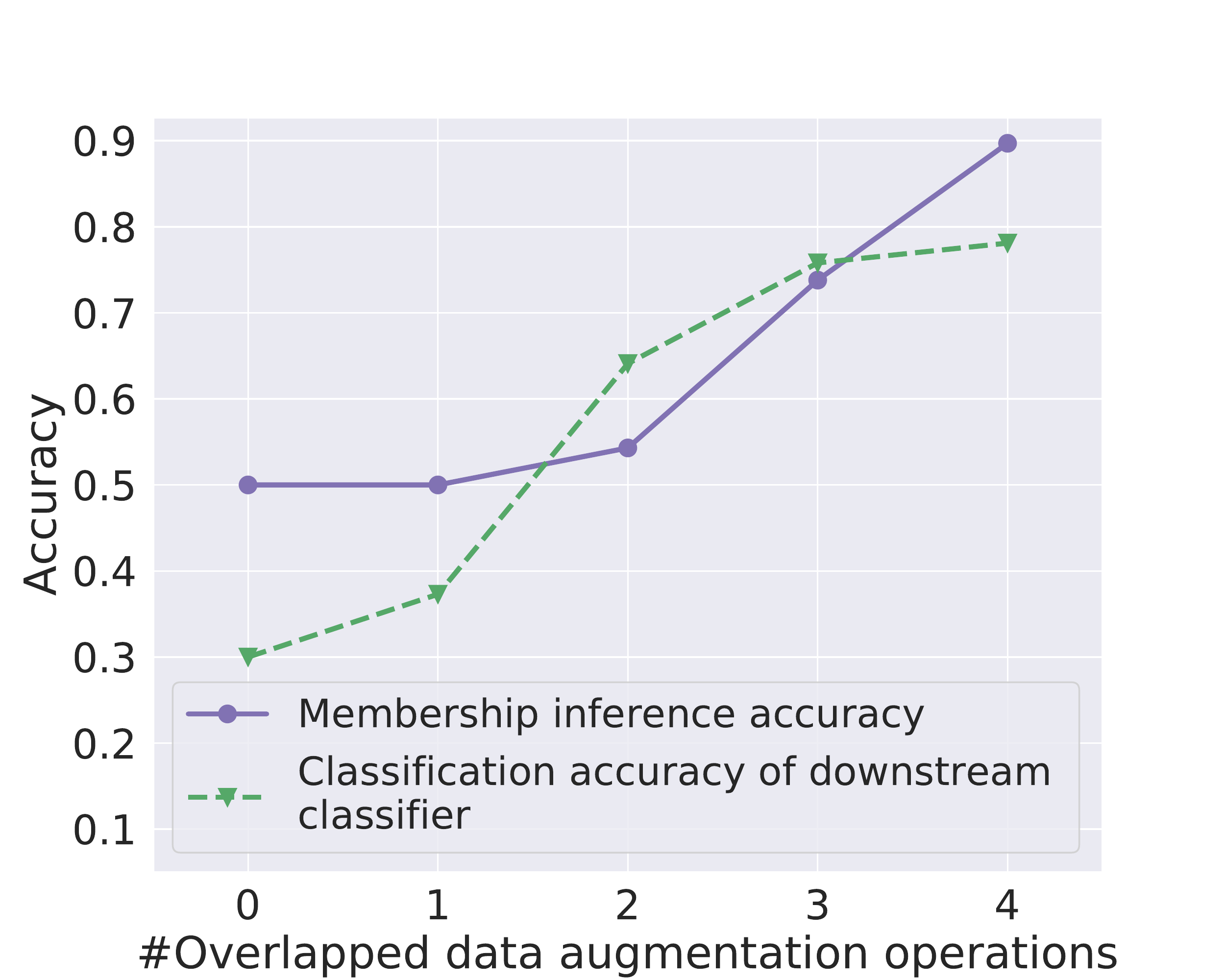}
\vspace{-3mm}
\caption{\SHF{Impact of  data augmentation, where the \SHF{method} is \SHF{EncoderMI}-V and the dataset is CIFAR10.} }
\vspace{-3mm}
\label{impact-data-aug}
\end{figure}

\myparatight{\SHF{Impact of data augmentation}} \SHF{The data augmentation operations an inferrer uses may be different from those used to pre-train the target encoder. In this experiment, we explicitly study the impact of data augmentation. We assume the inferrer uses a comprehensive list of four commonly used data augmentation operations, i.e., random grayscale, random resized crop, random horizontal flip, and color jitter.} \SHF{We gradually increase the number of overlapped data augmentation operations between the \SHF{inferrer}'s comprehensive list and the target encoder. In particular, the target encoder starts from only using data augmentation operation Gaussian blur, which is not in the \SHF{inferrer}’s list. Then, we add the \SHF{inferrer}'s data augmentation operations to the target encoder's pre-training module one by one in the following order: random grayscale, random resized crop, random horizontal flip, and color jitter. We calculate the membership inference accuracy and the classification accuracy of the downstream classifier for each target encoder, where the downstream classifier is built as in Baseline-A.   Figure~\ref{impact-data-aug} shows the results. We observe the number of overlapped data augmentation operations between the referrer and the target encoder controls a trade-off between membership inference accuracy and utility of the target encoder, i.e., the target encoder is more resistant against membership inference but the downstream classifier is also less accurate when the target encoder uses less data augmentation operations from the referrer's comprehensive list. 
}

\section{Applying Our Method to CLIP}
\label{attack_clip_section}
 CLIP~\cite{radford2021learning}  jointly pre-trains an image encoder and a text encoder on 400 million (image, text) pairs collected from the Internet. OpenAI has made the image encoder and text encoder publicly available. We view CLIP's image encoder with the ViT-B/32 architecture as the target encoder and apply our \SHF{EncoderMI} to infer its members. Specifically, given an input image, we aim to use  \SHF{EncoderMI} to infer whether it was used  by CLIP or not.  
 Next, we first introduce experimental setup and then present experimental results.

\subsection{Experimental Setup} 

\myparatight{Potential members and ground truth non-members} To evaluate our \SHF{EncoderMI} for CLIP's image encoder, we need an evaluation dataset consisting of both ground truth members and non-members. However, the pre-training dataset of CLIP is not released to the public. Therefore, we cannot obtain ground truth members of CLIP. However, we can collect some images that are \emph{potential members} and \emph{ground truth non-members} of the CLIP's pre-training dataset. Specifically, 
according to Radford et al.~\cite{radford2021learning},   the (image, text) pairs used to pre-train CLIP were collected from the Internet based on a set of 500,000 popular keywords. 
Therefore, we collect the following two evaluation datasets, each of which has 1,000 potential members and 1,000 ground truth non-members:
\begin{packeditemize}
    \item {\bf Google.} We use the class names of CIFAR100~\cite{krizhevsky2009learning} as keywords and use Google image search to collect images.  Appendix~\ref{words-google} shows the complete list of class names, e.g., ``clock'', ``house'', and ``bus''.  In particular, we use a publicly available tool\footnote{https://github.com/hardikvasa/google-images-download} to crawl images from Google search based on the keywords. We collected 10 images for each keyword, and thus we collected 1,000 images in total. We treat these images as \emph{potential members} as they were potentially also collected and used by CLIP.  To construct ground truth non-members, we further collected 2,000 images from Google search using the keywords. We randomly divided them into 1,000 pairs; and for each pair, we resized the two images  to  
the same size, and we concatenated them to form a new image, which results in 1,000 images in total. We treat these images as ground truth non-members of CLIP.

  \begin{figure}[!t]
\centering
\vspace{-5mm}
\subfloat[Google]{\includegraphics[width=0.25\textwidth]{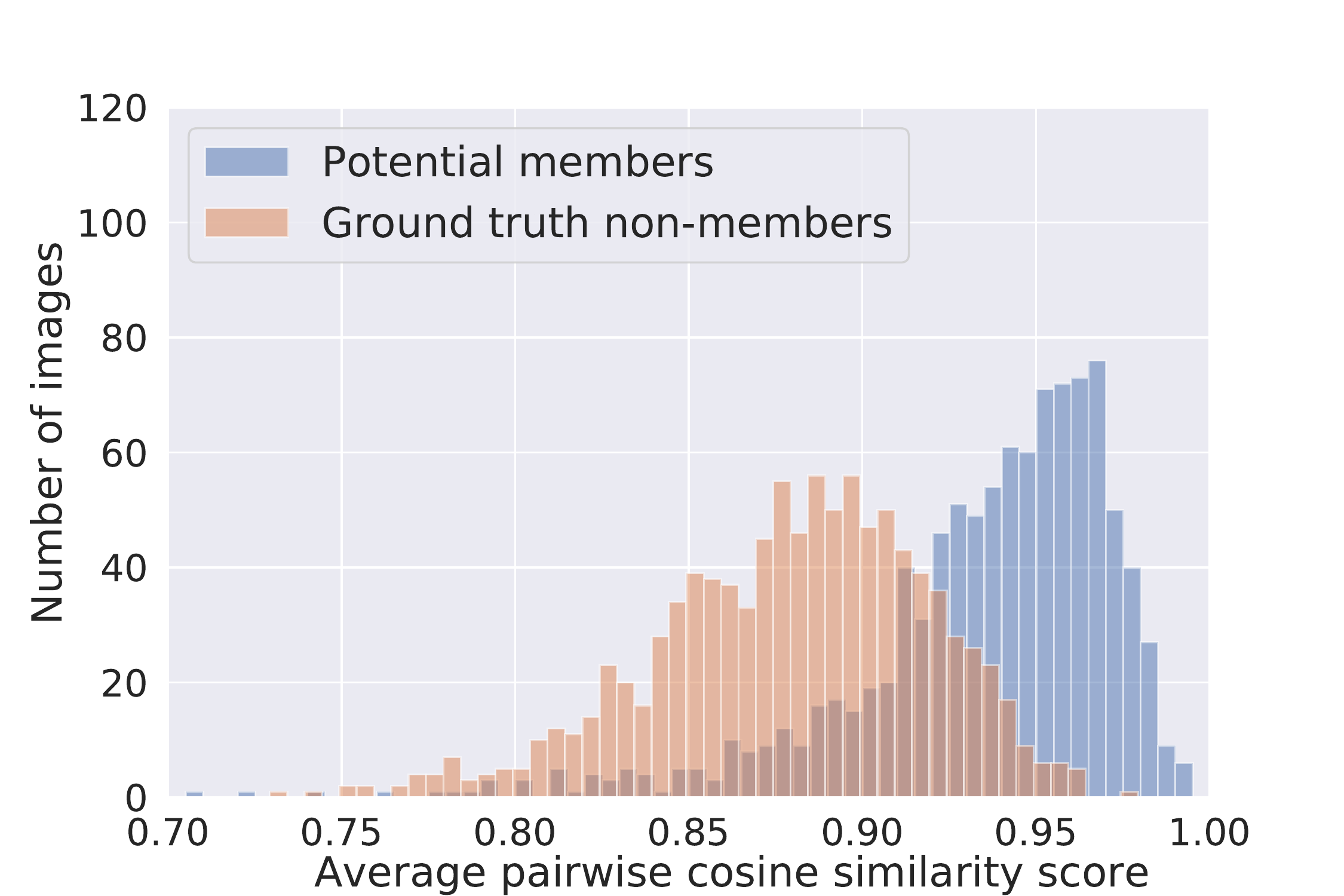}}
\subfloat[Flickr]{\includegraphics[width=0.25\textwidth]{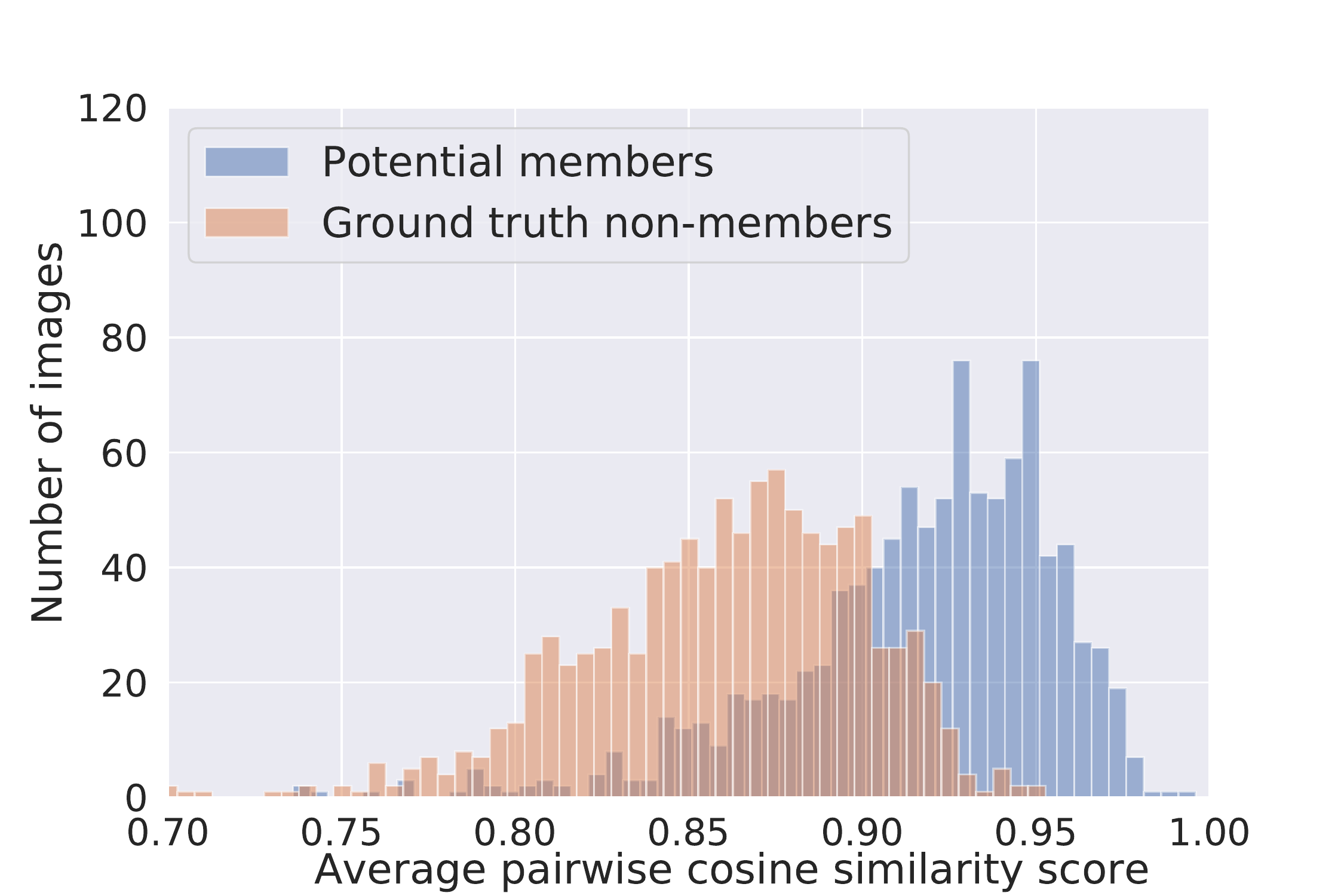}}
\vspace{-3mm}
\caption{Histograms of the average pairwise cosine similarity for potential members and ground truth non-members.}
\label{fig-google2}
\vspace{-5mm}
\end{figure}

    \item {\bf Flickr.} Similar to the above Google evaluation dataset, we collected an evaluation dataset from Flickr using the 100 keywords and a publicly available tool\footnote{https://stuvel.eu/software/flickrapi}. 
    Specifically, we collected 1,000 images as potential members. Moreover, we further collected 2,000 images and randomly paired them to be 1,000 images, which we treat as ground truth non-members. 
\end{packeditemize}

We acknowledge that some of the potential members may not be ground truth members of CLIP in both of our evaluation datasets. For each potential member and ground truth non-member, we resize them  as the input size of  CLIP, which is $224 \times 224$.

\begin{table}[!t]\renewcommand{\arraystretch}{1}
    \small
     \centering
          \caption{Accuracy, precision, and recall (\%) of \SHF{EncoderMI} for CLIP's image encoder.}
          \vspace{-2mm}
    \subfloat[Google]{
     \begin{tabular}{|c |c| c| c|c|}
     \hline
    \makecell{\SHF{Method}} & \makecell{Shadow dataset}  & Accuracy &Precision &Recall\\
     \hline
     \multirow{3}{*}{\makecell{\SHF{EncoderMI}-V}} & CIFAR10 &70.0 &64.5 &88.8 \cr\cline{2-5} 
     & STL10 &71.0 &65.4 &88.9 \cr\cline{2-5} 
     & Tiny-ImageNet & 67.6 & 62.0 &  90.7\cr\cline{1-5}  
     \multirow{3}{*}{\makecell{\SHF{EncoderMI}-S}} & CIFAR10 & 74.7 & 70.2 & 87.1 \cr\cline{2-5} 
     & STL10 & 74.6& 70.2& 86.8 \cr\cline{2-5} 
     & Tiny-ImageNet & 73.2& 68.1& 88.1\cr\cline{1-5} 
     \multirow{3}{*}{\makecell{\SHF{EncoderMI}-T}} & CIFAR10 & 70.3 & 64.2 & 89.7 \cr\cline{2-5} 
     & STL10 & 71.5 &65.6  & 90.0 \cr\cline{2-5} 
     & Tiny-ImageNet &66.3 & 62.4 & 90.1 \cr\cline{1-5}  
     \end{tabular}
     \label{tab: attack-clip-google}\hspace{2mm}}
     
     \subfloat[Flickr]{
          \begin{tabular}{|c |c| c| c|c|}
     \hline
    \makecell{\SHF{Method}} & \makecell{Shadow dataset}  & Accuracy &Precision &Recall\\
     \hline
     \multirow{3}{*}{\makecell{\SHF{EncoderMI}-V}} & CIFAR10 &74.9 &72.0 &81.5 \cr\cline{2-5} 
     & STL10 &73.5 &70.9 &79.3 \cr\cline{2-5}
     & Tiny-ImageNet & 74.7 & 73.1 &  78.2\cr\cline{1-5}  
     \multirow{3}{*}{\makecell{\SHF{EncoderMI}-S}} & CIFAR10 & 71.7 & 69.1 & 79.2 \cr\cline{2-5} 
     & STL10 & 72.7 & 71.4 & 76.4 \cr\cline{2-5} 
     & Tiny-ImageNet & 71.6 & 69.3& 79.0 \cr\cline{1-5} 
     \multirow{3}{*}{\makecell{\SHF{EncoderMI}-T}} & CIFAR10 & 73.9 & 68.8 & 80.1 \cr\cline{2-5} 
     & STL10 & 74.5 & 70.8 & 80.8 \cr\cline{2-5} 
     & Tiny-ImageNet &74.3 & 71.2 & 79.4 \cr\cline{1-5}  
     \end{tabular}
     \label{tab: attack-clip-flickr} 
}
\vspace{-5mm}
\label{tab: attack-clip}
 \end{table}

\myparatight{\SHF{Inference} classifiers} We assume the \SHF{inferrer} does not know the pre-training data distribution, encoder architecture, and training algorithm of CLIP, which is the most difficult scenario for our \SHF{EncoderMI}. 
We use the \SHF{inference} classifiers we built in our previous experiments in Section~\ref{evaluation-section}. Specifically, in our previous experiments, for each of our three \SHF{methods} (i.e., \SHF{EncoderMI}-V, \SHF{EncoderMI}-S, and \SHF{EncoderMI}-T) and each of the three shadow datasets (i.e., CIFAR10, STL10, and Tiny-ImageNet), we have built 8 \SHF{inference} classifiers corresponding to the 8 types of background knowledge. We use the \SHF{inference} classifiers corresponding to the background knowledge $\mathcal{B}=(\surd,\surd,\surd)$ in our previous experiments (i.e., the \SHF{inference} classifiers corresponding to the last rows of Table~\ref{tab: attack-results-cifar}, Table~\ref{tab: attack-results-stl}, and Table~\ref{tab: attack-results-tinyimagenet}) to \SHF{infer members of} CLIP.  
Given an input image, we create 10 augmented inputs and use the CLIP's image encoder to produce a feature vector for each of them. Then, we compute the 45 pairwise cosine similarity scores between the 10 feature vectors, which constitute the set of membership features for the input image. Our \SHF{inference} classifiers predict the membership status of the input image based on the membership features.

\subsection{Experimental Results}

\noindent
{\bf Cosine similarity score distribution for potential members and ground truth non-members:} Recall that, for each potential member or ground truth non-member,  \SHF{EncoderMI} constructs membership features consisting of 45 pairwise cosine similarity scores.  We compute the average of the 45 pairwise  cosine similarity scores for each potential member or ground truth non-member.  Figure~\ref{fig-google2} shows the histograms of the average pairwise cosine similarity scores for potential members and ground truth non-members in our two evaluation datasets.  
We observe that the potential members and ground truth non-members are statistically distinguishable with respect to the average pairwise cosine similarity scores. In particular, the potential members tend to have larger average pairwise cosine similarity scores than the ground truth non-members.

\noindent
{\bf \SHF{Effectiveness of EncoderMI}:} Table~\ref{tab: attack-clip} shows the accuracy, precision, and recall of the three variants of  \SHF{EncoderMI} based on different shadow datasets when applied to the CLIP's image encoder. The accuracy, precision, and recall are calculated based on the 1,000 potential members and 1,000 ground truth non-members in each evaluation dataset. First, we observe that  \SHF{EncoderMI} based on different shadow datasets achieves high accuracy, e.g., 0.66 -- 0.75. Our results imply that overfitting exists in real-world image encoders such as CLIP. Second,  \SHF{EncoderMI} achieves a higher recall than precision, which means that  \SHF{EncoderMI} predicts a large fraction of potential members as members. Third, our \SHF{EncoderMI} achieves higher  recall (or lower precision) on Google than Flickr. In other words, our \SHF{EncoderMI} predicts more inputs from Google image search as members. The reason is that, on average, the average pairwise cosine similarity score for inputs from  Google image search is larger than that for inputs from Flickr as shown in Figure~\ref{fig-google2}.

\section{Discussion on Countermeasures}

\noindent
{\bf Preventing overfitting via early stopping:}
Recall that our \SHF{EncoderMI} exploits the overfitting of a target encoder on its pre-training dataset. Note that the overfitting of a target encoder on its pre-training dataset is different from that of a classifier. For instance, when a classifier is overfitted to its training dataset, it may have different classification accuracies on its training dataset and testing dataset. Moreover, the confidence score vectors outputted by the classifier for its training dataset and testing dataset are also statistically distinguishable. Given an input, a target encoder outputs a feature vector for it. However, the feature vector itself does not capture the overfitting of the target encoder on the input. Instead, when the target encoder is overfitted to its pre-training dataset, it may output more similar feature vectors for the augmented versions of an input  in  the pre-training dataset. 

\begin{figure}[!t]
\centering
\vspace{-5mm}
\subfloat[Overfitting of the target encoder]{\includegraphics[width=0.25\textwidth]{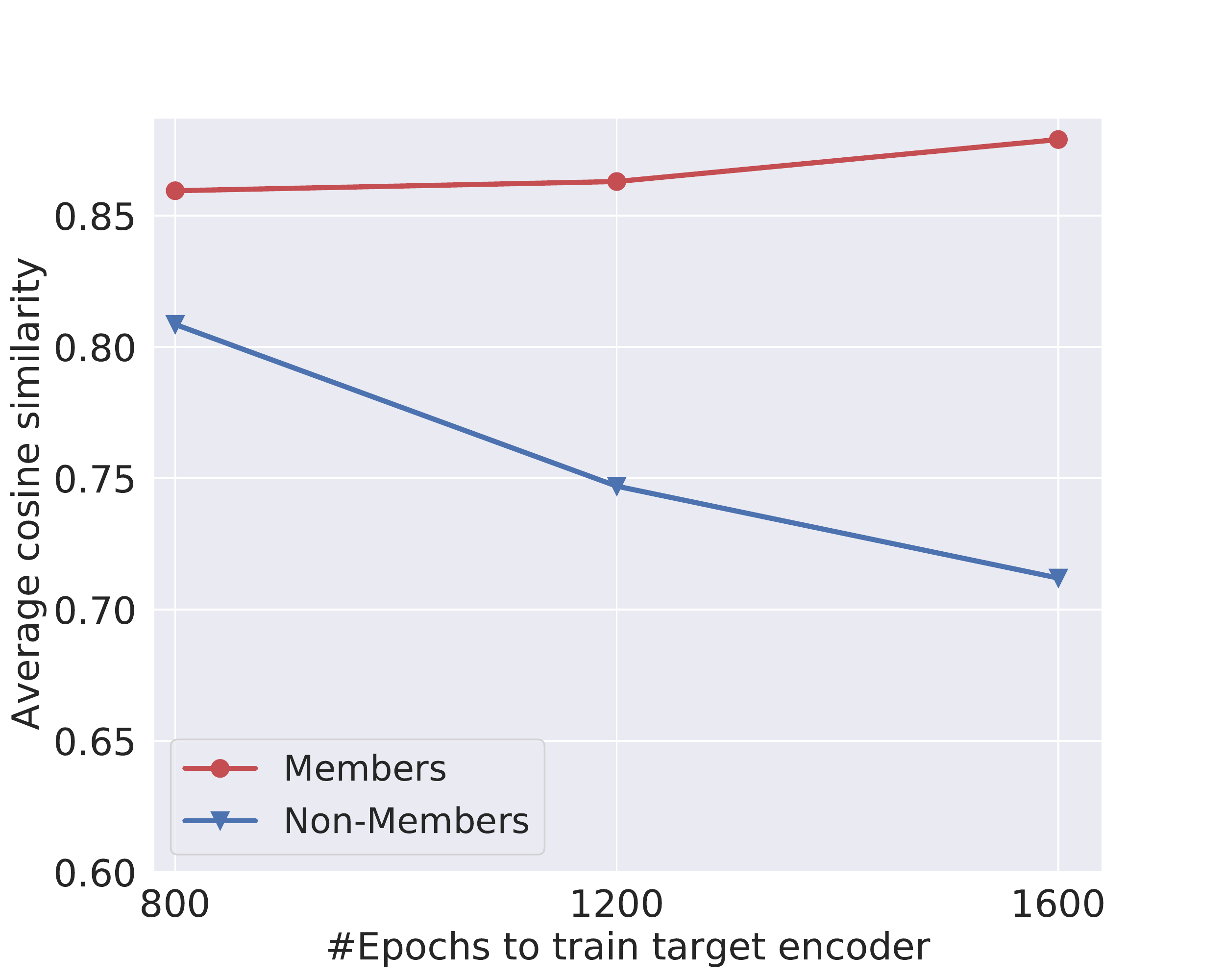}\label{overfitting_of_target_encoder}}
\subfloat[Inferability-utility tradeoff]{\includegraphics[width=0.25\textwidth]{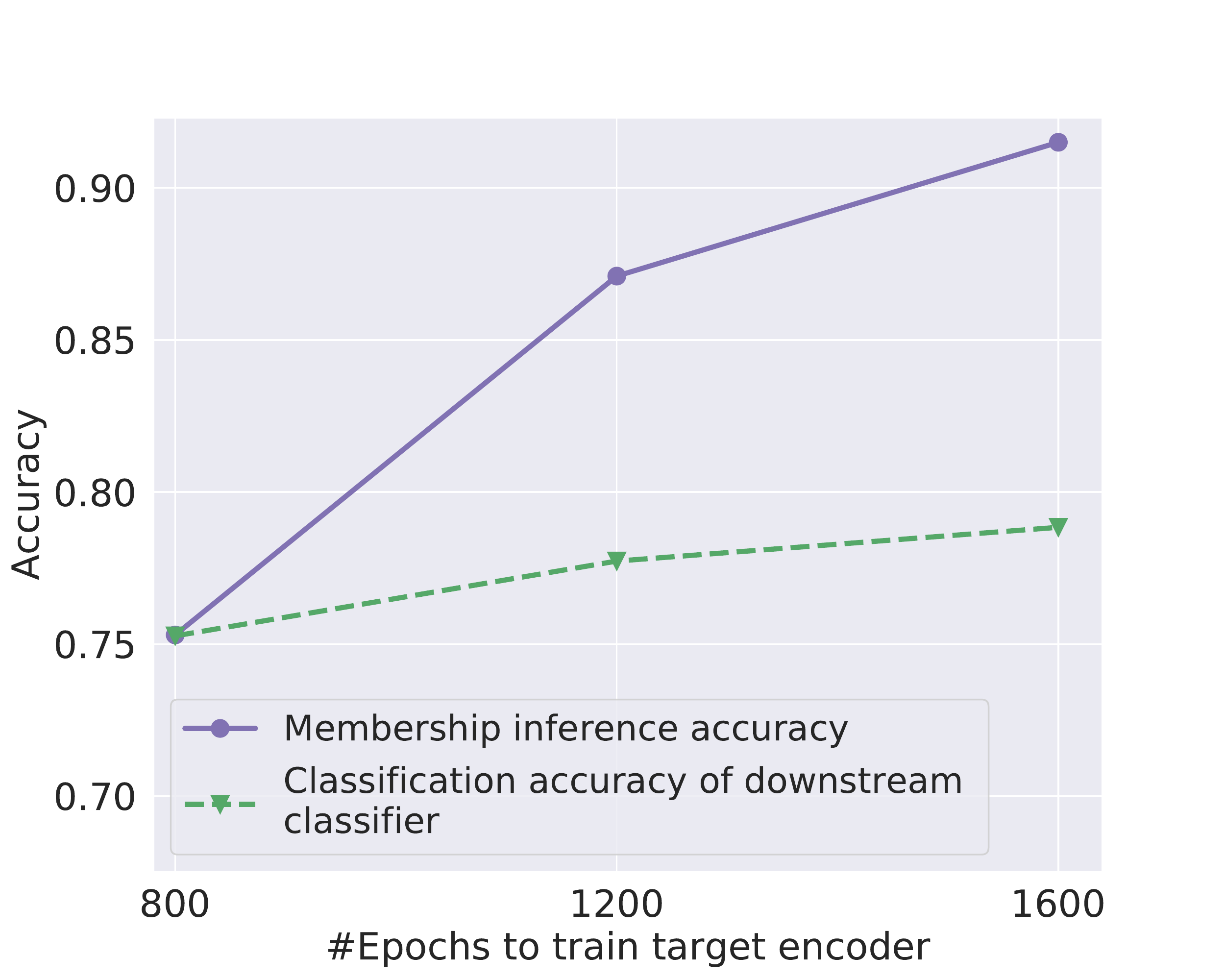}\label{privacy_utility_tradeoff}}
\vspace{-2mm}
\caption{(a)  The target encoder is  more overfitted to its pre-training data when it's pre-trained for more epochs. (b) Trade-off between membership inference accuracy and encoder utility for early stopping, where the method is \SHF{EncoderMI-V} and the dataset is CIFAR10.}
\vspace{-4mm}
\label{fig}
\end{figure}

We find that a target encoder is more overfitted to its pre-training dataset when it is trained for more epochs. 
Recall that the membership features for an input constructed by our \SHF{EncoderMI} consist of 45 pairwise cosine similarity scores. For each member or non-member of the target encoder,  we compute the average pairwise cosine similarity scores in its membership features, and we further compute the average pairwise cosine similarity scores over all members (or non-members). 
Figure~\ref{overfitting_of_target_encoder} shows the average pairwise cosine similarity scores for members and non-members of the target encoder as the number of pre-training epochs increases, where the pre-training dataset is based on CIFAR10.  
We observe the average pairwise cosine similarity increases (or decreases) for members (or non-members) as the number of epochs increases. In other words, the target encoder is more overfitted to its pre-training dataset when pre-training it for more epochs.

Our observation inspires us to counter \SHF{EncoderMI} via preventing overfitting through early stopping, i.e., pre-training a target encoder for less number of epochs.  
We evaluate the early stopping based countermeasure against our \SHF{EncoderMI}. In particular, we pre-train a target encoder on the pre-training dataset based on CIFAR10. After we pre-train the target encoder for some epochs, we calculate the accuracy of our \SHF{EncoderMI-V} under the background knowledge $\mathcal{B}=(\surd,\surd,\surd)$ and we also calculate the classification accuracy of a downstream classifier built based on the target encoder. Both the pre-training dataset and the downstream dataset are constructed based on CIFAR10 as we described in Section~\ref{evaluation-section}, and the classification accuracy of the downstream classifier is calculated based on the testing dataset of CIFAR10.   Figure~\ref{privacy_utility_tradeoff} shows
the membership inference accuracy of our \SHF{EncoderMI-V} and the classification accuracy of the downstream classifier as we pre-train the target encoder for more epochs. We observe that the early stopping based defense achieves a trade-off, i.e., it decreases the membership inference accuracy but also reduces the classification accuracy of the downstream classifier. We note that Song et al.~\cite{song2020systematic} found that early stopping  outperforms other overfitting-prevention countermeasures against membership inference  to classifiers, and they also observed a trade-off between membership inference accuracy and classifier utility for early stopping.

\noindent
{\bf Pre-training with differential privacy:}
Differential privacy~\cite{dwork2006calibrating,shokri2015privacy,abadi2016deep,jayaraman2019evaluating,nasr2021adversary} can provide formal membership privacy guarantees for each input in the training dataset of a machine learning model. Many differentially private learning algorithms have been proposed. These algorithms add noise to the training data~\cite{duchi2013local}, the objective function~\cite{iyengar2019towards,jayaraman2019evaluating}, or the gradient computed by (stochastic) gradient descent during the learning process~\cite{shokri2015privacy,abadi2016deep,jayaraman2019evaluating}. For instance, Abadi et al.~\cite{abadi2016deep} proposed differentially private stochastic gradient descent (DP-SGD) which adds random Gaussian noise to the gradient computed by stochastic gradient descent. It would be interesting future work to generalize these differentially private  learning algorithms to contrastive learning. In particular, when the pre-training dataset changes by one input, the encoder learnt by a differentially private contrastive learning algorithm does not change much. However, differential privacy may also incur  large utility loss for the encoder, i.e., the downstream classifiers built based on a differentially private encoder may have much lower classification accuracy. 

\noindent
{\bf Adversarial learning:}
Adversarial learning based countermeasures~\cite{nasr2018machine,jia2019memguard} have been studied to mitigate  membership inference to classifiers, which were inspired by adversarial learning based countermeasures against attribute inference attacks~\cite{jia2018attriguard}. 
For instance, Nasr et al.~\cite{nasr2018machine} proposed to add an adversarial regularization term to the loss function when training a target classifier, where the adversarial regularization term models a membership inference method's accuracy. Jia et al.~\cite{jia2019memguard} proposed MemGuard which does not modify the training process, but adds carefully crafted perturbation to the confidence score vector outputted by the target classifier for each input. Specifically, the idea is to turn the perturbed confidence score vector to be an adversarial example to the \SHF{inference} classifiers, which make random membership inference based on the perturbed confidence score vector. It would be interesting future work to extend these countermeasures to pre-trained encoders. For instance, we may capture our \SHF{EncoderMI}'s accuracy as an adversarial regularization term and add it to the contrastive loss when pre-training an encoder using contrastive learning; we may also add carefully crafted perturbation to the feature vector outputted by an encoder for each (augmented) input such that the set of membership features constructed by our \SHF{EncoderMI} for an input becomes an adversarial example to the \SHF{inference} classifiers,  which make random membership inference based on the perturbed membership features. A key challenge is how to find such perturbation to each feature vector since the membership features depend on the pairwise similarity between a set of feature vectors corresponding to the augmented versions of an input.

\section{Related Work}
\label{related_work}

\noindent
{\bf Membership inference:} In membership inference against machine learning classifiers~\cite{shokri2017membership,yeom2018privacy,salem2018ml,sablayrolles2019white,nasr2019comprehensive,song2019privacy,hagestedt2019mbeacon,choo2020label,hagestedt2020membership,hui2021practical,nasr2019comprehensive,li2021label,li2020membership,song2020systematic,he2021quantifying,li2021label}, an \SHF{inferrer} aims to infer whether an input is in the training dataset of a classifier (called \emph{target classifier}).  
For instance, in the methods proposed by Shokri et al.~\cite{shokri2017membership}, an \SHF{inferrer}  first trains shadow classifiers to mimic the behaviors of the target classifier. Given the shadow classifiers whose ground truth members and non-members are known to the \SHF{inferrer}, the \SHF{inferrer}  trains \SHF{inference} classifiers, which are then applied to infer members of the target classifier. Salem et al.~\cite{salem2018ml} further improved these methods by relaxing the assumptions about the \SHF{inferrer}. Hui et al.~\cite{hui2021practical} proposed blind membership inference methods that do not require training shadow classifiers. Concurrent to our work, He et al.~\cite{he2021quantifying} also studied membership inference against contrastive learning. They assume the pre-training data and downstream data are the same. Specifically, given a labeled training dataset, they first use contrastive learning to pre-train an  encoder and then use it to fine-tune a classifier on the labeled training dataset. 
They try to infer whether an input is in the labeled training dataset via applying existing methods~\cite{shokri2017membership,salem2018ml} to the fine-tuned classifier. 

Our methods are different from these ones as they were designed to infer  members of a \emph{classifier} while our methods aim to infer members of an encoder pre-trained by \SHF{contrastive} learning. 
Our experimental results show that these methods achieve accuracy close to random guessing when applied to infer members of an encoder. The reason is that  
the confidence score vector outputted by a classifier can capture whether the classifier is overfitted for an input, while the feature vector itself outputted by an encoder does not capture whether the encoder is overfitted for an input. The similarity scores between the feature vectors of the augmented versions of an input capture whether the encoder is overfitted for the input, and our methods leverage such similarity scores to infer the membership status of the input.

Existing membership inference methods for pre-trained models~\cite{song2020information,carlini2020extracting} focused on the natural language domain. For instance, Carlini et al.~\cite{carlini2020extracting} proposed membership inference methods for GPT-2~\cite{radford2019language}, which is a pre-trained language model, and they further leveraged the membership inference methods to reconstruct the training data of GPT-2. Specifically, they first reconstructed some candidate texts and then applied a membership inference method to determine the membership status of each candidate text.    
To the best of our knowledge, no prior work has studied membership inference  for encoders in the image domain.

\SHF{Prior work~\cite{zou2020privacy,hidano2020transmia} also studied membership inference against transfer learning. 
For instance, Hidano et al.~\cite{hidano2020transmia} assume a white-box access to the transferred part of the teacher model while Zou et al.~\cite{zou2020privacy} leverage the posterior of the teacher model. Our work is different from these, because pre-training an image encoder is different from training a teacher model since the former uses contrastive learning on unlabeled data while the latter uses the standard supervised learning on labeled data. }

\noindent
{\bf Countermeasures against membership inference:} Many countermeasures~\cite{shokri2017membership,salem2018ml,nasr2018machine,chen2018differentially,jia2019memguard,li2020membership,song2020systematic} were proposed to counter membership inference for classifiers. The first category of countermeasures~\cite{shokri2017membership,salem2018ml,song2020systematic}  try to prevent overfitting when training classifiers, e.g., standard $L_2$ regularization~\cite{shokri2017membership}, dropout~\cite{salem2018ml}, and early stopping~\cite{song2020systematic}. The second category of countermeasures~\cite{shokri2015privacy,abadi2016deep} are based on differential privacy~\cite{dwork2006calibrating}, which often incur large utility loss for the learnt machine learning classifiers. The third category of countermeasures leverage adversarial learning, e.g., adversarial regularization~\cite{nasr2018machine} and MemGuard~\cite{jia2019memguard}. We  explored an early stopping based countermeasure against our \SHF{EncoderMI}. Our results show that such countermeasure achieves a trade-off between membership inference accuracy and  utility of an encoder.

\noindent
{\bf Contrastive learning:}
Contrastive learning~\cite{hadsell2006dimensionality,ranzato2007unsupervised,vincent2008extracting,dosovitskiy2015discriminative,oord2018representation,he2020momentum,chen2020simple} aims to pre-train image encoders on unlabeled data via exploiting the supervisory signals in the unlabeled data itself.   
The unlabeled data could be unlabeled images or (image, text) pairs. The pre-trained  encoders can be used  for many downstream tasks. The key idea of contrastive learning is to pre-train an image encoder such that it outputs similar feature vectors for a pair of augmented inputs created from the same input image and outputs dissimilar feature vectors for a pair of augmented inputs created from different input images. Examples of contrastive learning methods include MoCo~\cite{he2020momentum}, SimCLR~\cite{chen2020simple}, and CLIP~\cite{radford2021learning}, which we discussed in Section~\ref{sec:background}. We note that Jia et al.~\cite{jia2021badencoder} proposed BadEncoder, which embeds backdoors into a pre-trained image encoder such that multiple downstream classifiers built based on the backdoored encoder inherit the backdoor behavior simultaneously.

\section{Conclusion and Future Work}
\SHF{In this work, we propose the first membership inference method against image encoders pre-trained by contrastive learning. Our method exploits the overfitting of an image encoder, i.e., it produces more similar feature vectors for two augmented versions of the same input. Our experimental results on image encoders pre-trained on multiple datasets by ourselves as well as a real-world image encoder show that our method can achieve high accuracy, precision, and recall. Moreover, we also find that  an early stopping based countermeasure achieves a trade-off between membership inference accuracy and encoder utility.   }

\SHF{Interesting future work includes 1) extending our method to the white-box settings in which the inferrer has access to the parameters of the target encoder, 2) extending our method to infer the membership of an (image, text) pair,  3) developing new countermeasures against our method, and 4) exploring other privacy/confidentiality risks of pre-trained image encoders such as stealing their parameters~\cite{tramer2016stealing} and hyperparameters (e.g., encoder architecture)~\cite{wang2018stealing}.}

\myparatight{\SHF{Acknowledgements}} \SHF{ We thank the anonymous reviewers and our shepherd Reza Shokri for constructive comments. This work was supported by NSF under Grant No. 1937786.}

\bibliographystyle{ACM-Reference-Format}
\bibliography{refs}

\appendix

\begin{table*}[htb]\renewcommand{\arraystretch}{1}
     \centering
     \small
     \setlength{\tabcolsep}{1mm}
     \caption{Average accuracy, precision, and recall (\%) of our \SHF{methods} for the target encoder pre-trained on STL10 dataset. $\surd$ (or $ \times$) means the \SHF{inferrer} has
  (or does not have) access to the corresponding background knowledge of the target encoder. The numbers in parenthesis are standard deviations in 5 trials.}
  \vspace{-2mm}
     \begin{tabular}{|c|c|c|c|c|c|c|c|c|c|c|c|}
     \hline
     \multirow{3}{*}{\makecell{Pre-training\\data distribution}} & \multirow{3}{*}{\makecell{Encoder\\architecture}} & \multirow{3}{*}{\makecell{Training\\ algorithm}} &\multicolumn{3}{c|}{Accuracy} &\multicolumn{3}{c|}{Precision} &\multicolumn{3}{c|}{Recall} \cr\cline{4-12}  
      & & & \makecell{ \SHF{Encod-}\\ \SHF{erMI}-V} & \makecell{ \SHF{Encod-}\\ \SHF{erMI}-S} & \makecell{ \SHF{Encod-}\\ \SHF{erMI}-T} &  \makecell{ \SHF{Encod-}\\ \SHF{erMI}-V} & \makecell{ \SHF{Encod-}\\ \SHF{erMI}-S} & \makecell{ \SHF{Encod-}\\ \SHF{erMI}-T} 	&  \makecell{ \SHF{Encod-}\\ \SHF{erMI}-V} &\makecell{ \SHF{Encod-}\\ \SHF{erMI}-S} & \makecell{ \SHF{Encod-}\\ \SHF{erMI}-T}\\ \hline
           $\times$ & $\times$ & $\times$ & 		81.9 (1.94)  & 80.4 (1.79) 	& 81.3 (1.74)		& 	79.1 (2.35)  & 77.2	(2.04)	& 78.4 (1.87)		& 	90.2 (2.79)  & 92.0 (2.61) 	& 90.3 (2.03)		\\ \hline
           $\surd$  & $\times$ & $\times$ & 		83.8 (1.81)  & 83.4 (1.79) 	& 82.4 (1.72)		& 	81.2 (2.38)  & 79.8 (1.49)	& 80.1 (1.79)		& 	91.2 (2.32)  & 92.3 (2.41) & 91.2 (1.32)		\\ \hline
           $\times$ & $\surd$  & $\times$ & 		84.5 (1.72)  & 83.1 (1.62) 	& 82.6 (1.42)		& 	81.1 (1.87)  & 79.3 (1.84) 	& 78.9 (1.32)		& 	92.6 (1.26)  & 92.7 (1.35) 	& 92.6 (0.97)		\\ \hline
           $\times$ & $\times$ & $\surd$  & 		86.9 (1.04)  & 85.6 (0.98) 	& 85.6 (0.83)		& 	82.5 (0.88)  & 81.2 (0.79)		& 81.4 (0.73)		& 	96.4 (1.32)  & 93.6 (1.05)	 & 95.0 (0.94)	 \\ \hline
           $\surd$  & $\surd$	 & $\times$ & 		84.7 (1.53)  & 83.2 (1.46) 	& 82.7	 (1.24)	& 	81.3 (1.47)  & 	79.4 (1.29)	& 78.9 (1.03)		& 	92.9 (2.12)  & 91.7 (1.97)		& 91.6 (1.61)	 \\ \hline
           $\surd$  & $\times$ & $\surd$  & 		87.1 (0.69)  & 85.1 (0.69) 	& 86.1 (0.58)		& 	82.0 (0.71)  & 81.2 (0.67)		& 81.7 (0.61)		& 	97.1 (0.93)  & 96.2 (0.97) & 97.3 (0.86)\\ \hline
           $\times$ & $\surd$  & $\surd$  & 		90.0 (0.57)  & 85.8 (0.59) 	& 87.4 (0.44)		& 	87.0 (0.73)  & 84.8 (0.91)		& 83.6 (0.82)		& 	94.1 (0.99)  & 90.1 (1.04) 	& 93.2 (0.85)		\\ \hline
     	   $\surd$  & $\surd$  & $\surd$  & 		90.1 (0.54)  & 86.2 (0.66) 	& 88.7	 (0.49)	& 	87.2 (0.67)  & 83.1	(0.59)	& 83.2 (0.51)		& 	94.1 (1.04)  & 92.2 (0.92)	& 95.1 (1.01)		\\ \hline
     \end{tabular} 
     \label{tab: attack-results-stl}
 \end{table*}
 
   \begin{table*}[htb]\renewcommand{\arraystretch}{1}
     \centering
     \small
     \setlength{\tabcolsep}{1mm}
     \caption{Average accuracy, precision, and recall (\%) of our  \SHF{methods} for the target encoder pre-trained on Tiny-ImageNet dataset. $\surd$ (or $ \times$) means the \SHF{inferrer} has
  (or does not have) access to the corresponding background knowledge of the target encoder. The numbers in parenthesis are standard deviations in 5 trials.}
  \vspace{-2mm}
     \begin{tabular}{|c|c|c|c|c|c|c|c|c|c|c|c|}
     \hline
     \multirow{3}{*}{\makecell{Pre-training\\data distribution}} & \multirow{3}{*}{\makecell{Encoder\\architecture}} & \multirow{3}{*}{\makecell{Training\\ algorithm}} &\multicolumn{3}{c|}{Accuracy} &\multicolumn{3}{c|}{Precision} &\multicolumn{3}{c|}{Recall} \cr\cline{4-12}  
      & & & \makecell{ \SHF{Encod-}\\ \SHF{erMI}-V} & \makecell{ \SHF{Encod-}\\ \SHF{erMI}-S} & \makecell{ \SHF{Encod-}\\ \SHF{erMI}-T} &  \makecell{ \SHF{Encod-}\\ \SHF{erMI}-V} & \makecell{ \SHF{Encod-}\\ \SHF{erMI}-S} & \makecell{ \SHF{Encod-}\\ \SHF{erMI}-T} 	&  \makecell{ \SHF{Encod-}\\ \SHF{erMI}-V} &\makecell{ \SHF{Encod-}\\ \SHF{erMI}-S} & \makecell{ \SHF{Encod-}\\ \SHF{erMI}-T}\\ \hline
           $\times$ & $\times$ & $\times$ & 	88.7 (1.81)   &  84.9 (1.73) & 85.3 (1.67)	& 86.0 (1.98)	  & 81.5 (2.03)		& 81.8 (1.74)	& 90.1 (1.96)	  & 95.3 (1.67)	& 95.9 (1.44)\\  	 \hline
           $\surd$  & $\times$ & $\times$ &  	93.0 (1.74)  & 88.2 (1.68)	& 90.0 (1.44) 	& 	90.1 (1.39)  & 85.4 (1.45)		& 86.8 (1.23)	& 	97.8 (1.26)  & 93.2 (1.22)		& 97.1 (1.11)\\    \hline
           $\times$ & $\surd$  & $\times$ & 	89.1 (1.63)  & 86.4 (1.64) & 85.7 (1.29)	& 	83.3 (1.88) & 84.0 (1.84)		& 80.1 (1.63)		& 	96.3 (1.22)  & 91.1 (1.29)		& 96.1 (1.08)\\   \hline
           $\times$ & $\times$ & $\surd$  &	 	94.1 (1.07) & 91.3 (1.03) 	& 94.1 (0.91)	& 	90.7 (0.88)  & 90.3 (0.87) 	& 93.5 (0.79)	& 	97.4 (0.92)  & 91.3 (1.22)	& 95.6 (0.93)\\   \hline
           $\surd$  & $\surd$	 & $\times$ & 	94.4 (1.38)  & 90.4 (1.33) & 91.5 (1.26)	& 	97.4 (0.96)  & 94.1 (0.91)		& 93.8 (0.91)	& 	90.8 (1.46)  & 87.1 (1.37) 	& 89.4 (1.22)\\   \hline
           $\surd$  & $\times$ & $\surd$  & 	96.1 (0.67)  & 91.6 (0.69) & 94.1 (0.54)	& 	93.8 (0.73)  & 90.4 (0.68)	& 94.2 (0.62)		& 	97.6 (0.99)  & 92.3 (1.02)		& 95.7 (0.88)\\   \hline
           $\times$ & $\surd$  & $\surd$  & 	94.5 (0.59)  & 91.8 (0.56)	& 92.0 (0.53)	& 	92.3 (0.93)  & 94.4 (0.91)	& 94.1	(0.86)	& 	96.7 (0.92)  & 90.6 (0.79) 	& 92.7 (0.77) \\  \hline
     	   $\surd$  & $\surd$  & $\surd$  & 	96.5 (0.51)  & 92.0 (0.47)	& 94.3 (0.43)	& 	96.6 (0.72)  & 92.9 (0.59)		& 94.9 (0.57)		& 	97.0 (0.93)  & 92.4 (0.89) 	& 93.2 (0.91)\\  \hline
     \end{tabular}
     \label{tab: attack-results-tinyimagenet}
 \end{table*}

\begin{table}[!t]\renewcommand{\arraystretch}{1.2}
     \centering
     \caption{ \SHF{Accuracy, precision, and recall (\%) of Baseline-E with 3×1 patches and 3×5 patches.}}
     \setlength{\tabcolsep}{1mm}
     \vspace{-2mm}
     \SHF{
\subfloat[3 $\times$ 1 Patches]{
          \begin{tabular}{|c|c|c|c|}
     \hline
     Pre-training dataset & Accuracy  &Precision   &Recall  \\
     \hline
     CIFAR10 &  60.2&	57.4&	79.3
  \\ \hline
     STL10 &64.1&	63.7&	65.7
\\  \hline
     Tiny-ImageNet  &65.8&	68.2&	59.5
\\  
     \hline
     \end{tabular}
      \label{tab: 3x1}
     }
     }          
     \\
          \SHF{
\subfloat[3 $\times$ 5 Patches]{
          \begin{tabular}{|c|c|c|c|}
     \hline
     Pre-training dataset & Accuracy  &Precision   &Recall  \\
     \hline
     CIFAR10  &50.8	&50.6&	62.9
  \\ \hline
     STL10 &55.8&	55.5&	59.3
  \\  \hline
     Tiny-ImageNet  &52.7&	52.4&	58.4
\\  
     \hline
     \end{tabular}
      \label{tab: baseline-D}
     } \\
     }
     \label{tab: baseline-e-other}
     \vspace{-7mm}
 \end{table}

 \section{Keywords for Google Search and Flickr Cralwer}
\label{words-google}
The keywords  are the 100 class names in CIFAR100: beaver, dolphin, otter, seal, whale, aquarium fish, flatfish, ray, shark, trout, orchids, poppies, roses, sunflowers, tulips, bottles, bowls, cans, cups, plates, apples, mushrooms, oranges, pears, sweet peppers, clock, computer keyboard, lamp, telephone, television, bed, chair, couch, table, wardrobe, bee, beetle, butterfly, caterpillar, cockroach, bear, leopard, lion, tiger, wolf, bridge, castle, house, road, skyscraper, cloud, forest, mountain, plain, sea, camel, cattle, chimpanzee, elephant, kangaroo, fox, porcupine, possum, raccoon, skunk, crab, lobster, snail, spider, worm, baby, boy, girl, man, woman, crocodile, dinosaur, lizard, snake, turtle, hamster, mouse, rabbit, shrew, squirrel, maple, oak, palm, pine, willow, bicycle, bus, motorcycle, pickup truck, train, lawn-mower, rocket, streetcar, tank, tractor.

\end{document}